\documentclass[11pt]{article}
\usepackage{setspace}   
\usepackage{jheppub}
\usepackage{amsfonts}
\allowdisplaybreaks[1]
\usepackage[font=small,labelfont=bf]{caption} 
\usepackage{hyperref}

\newcommand{\R}{{\mathbb R}}
\newcommand{\C}{{\mathbb C}}
\newcommand{\eq}[1]{Eq.~(\ref{#1})}

\newcommand{\cN}{{\mathcal{N}}}

\newcommand{\one}{{\rm 1\kern -.9mm l}}

\newcommand{\ft}[2]{{\textstyle\frac{#1}{#2}}}

\def\XXint#1#2#3{{\setbox0=\hbox{$#1{#2#3}{\int}$}
     \vcenter{\hbox{$#2#3$}}\kern-.5\wd0}}

\def\nn{\nonumber}

\newcommand{\beqas}{\begin{eqnarray*}}
\newcommand{\eeqas}{\end{eqnarray*}}
\def\eeq{\end{equation}}
\def\be{\begin{equation}}
\def\ee{\end{equation}}
\def\bea{\begin{eqnarray}}
\def\eea{\end{eqnarray}}
\def\eq{\begin{equation}}
\def\eqe{\end{equation}}
\def\eqa{\begin{eqnarray}}
\def\eqae{\end{eqnarray}}
\def\nn{\nonumber}

\usepackage{xcolor}
\usepackage{xparse}
\usepackage{mleftright}
\usepackage{xspace}
\usepackage{dsfont}
\usepackage{multirow}
\usepackage{subfig}
\usepackage{tikz}
\usetikzlibrary{arrows.meta}

\newcommand{\unit}{\mathds{1}}

\newcommand{\CF}{\mathcal{F}}
\newcommand{\CG}{\mathcal{G}}

\newcommand{\CJ}{\mathcal{J}}

\newcommand{\CN}{\mathcal{N}}
\newcommand{\CO}{\mathcal{O}}
\newcommand{\COb}{{\overbar{\mathcal{O}}}}

\newcommand{\CV}{\mathcal{V}}

\newcommand{\CY}{\mathcal{Y}}

\newcommand{\phib}{{\bar{\phi}}}
\newcommand{\zb}{{\bar{z}}}
\newcommand{\jb}{{\bar{\jmath}}}

\newcommand{\overbar}[1]{\mkern 3.5mu\overline{\mkern-3.5mu#1\mkern-1.5mu}\mkern 1.5mu}

\newcommand{\lsp}{\hspace{1pt}}
\newcommand{\llsp}{\hspace{0.5pt}}
\newcommand{\lnsp}{\hspace{-0.8pt}}

\makeatletter
\def\@fpheader{}
\makeatother

\NewDocumentCommand\eqn{mo}{%
  \IfNoValueTF{#2}
     {\[ #1 \]}
     {\begin{equation}\label{#2} #1 \end{equation} \expandafter\newcommand\csname #2\endcsname{\eqref{#2}\xspace}\ignorespaces}
}
\NewDocumentCommand\eqna{mo}{%
  \IfNoValueTF{#2}
    {\begin{align*} #1 \end{align*}}
    {\begin{equation}\label{#2}\begin{aligned} #1 \end{aligned}\end{equation} \expandafter\def\csname #2\endcsname{\eqref{#2}\xspace}\ignorespaces}
}

\NewDocumentCommand\twoseqn{momoo}{%
    \IfNoValueTF{#5}
       {\begin{subequations}\begin{align} #1\label{#2} \\ #3 \label{#4}  \end{align}\end{subequations} \expandafter\def\csname #2\endcsname{\eqref{#2}\xspace}\ignorespaces \expandafter\def\csname #4\endcsname{\eqref{#4}\xspace}\ignorespaces}
       {\begin{subequations}\label{#5}\begin{align} #1\label{#2} \\ #3 \label{#4}  \end{align}\end{subequations} \expandafter\def\csname #5\endcsname{\eqref{#5}\xspace}\ignorespaces \expandafter\def\csname #2\endcsname{\eqref{#2}\xspace}\ignorespaces \expandafter\def\csname #4\endcsname{\eqref{#4}\xspace}\ignorespaces}
}
\NewDocumentCommand\threeseqn{momomoo}{%
   \IfNoValueTF{#7}
     {\begin{subequations}\begin{align} #1\label{#2} \\ #3 \label{#4} \\ #5 \label{#6} \end{align}\end{subequations} \expandafter\def\csname #2\endcsname{\eqref{#2}\xspace}\ignorespaces \expandafter\def\csname #4\endcsname{\eqref{#4}\xspace}\ignorespaces \expandafter\def\csname #6\endcsname{\eqref{#6}\xspace}\ignorespaces}
     {\begin{subequations}\label{#7}\begin{align} #1\label{#2} \\ #3 \label{#4} \\ #5 \label{#6} \end{align}\end{subequations} \expandafter\def\csname #7\endcsname{\eqref{#7}\xspace}\ignorespaces \expandafter\def\csname #2\endcsname{\eqref{#2}\xspace}\ignorespaces \expandafter\def\csname #4\endcsname{\eqref{#4}\xspace}\ignorespaces \expandafter\def\csname #6\endcsname{\eqref{#6}\xspace}\ignorespaces}
}
\NewDocumentCommand\fourseqn{momomomoo}{%
   \IfNoValueTF{#9}
     {\begin{subequations}\begin{align} #1\label{#2} \\ #3 \label{#4} \\ #5 \label{#6} \\ #7\label{#8} \end{align}\end{subequations} \expandafter\def\csname #2\endcsname{\eqref{#2}\xspace}\ignorespaces \expandafter\def\csname #4\endcsname{\eqref{#4}\xspace}\ignorespaces \expandafter\def\csname #6\endcsname{\eqref{#6}\xspace}\ignorespaces \expandafter\def\csname #8\endcsname{\eqref{#8}\xspace}\ignorespaces}
     {\begin{subequations}\label{#9}\begin{align} #1\label{#2} \\ #3 \label{#4} \\ #5 \label{#6} \\ #7\label{#8} \end{align}\end{subequations} \expandafter\def\csname #9\endcsname{\eqref{#9}\xspace}\ignorespaces \expandafter\def\csname #2\endcsname{\eqref{#2}\xspace}\ignorespaces \expandafter\def\csname #4\endcsname{\eqref{#4}\xspace}\ignorespaces \expandafter\def\csname #6\endcsname{\eqref{#6}\xspace}\ignorespaces \expandafter\def\csname #8\endcsname{\eqref{#8}\xspace}\ignorespaces}
}

\allowdisplaybreaks

\graphicspath{{figures/}}
\newcommand{\includegraphicsWlabel}[4][]{\begin{tikzpicture}\node[label={[rotate=90, anchor=south]left:\footnotesize{\hspace{1em}#3}}, label={below:\footnotesize{\hspace{1.5em}#4}}] at (0,0) {\includegraphics[#1]{#2}};\end{tikzpicture}}

\title{\boldmath  OPE coefficients in Argyres-Douglas theories}
\author[1]{A. Bissi,}
\author[2]{F. Fucito,}
\author[1]{A. Manenti,}
\author[2]{J.F. Morales,}
\author[2]{and R. Savelli}
\affiliation[1]{
Department of Physics and Astronomy\\
Uppsala University, Box 516, SE-751 20 Uppsala, Sweden
}
\affiliation[2]{
I.N.F.N -- sezione di Roma Tor Vergata and Dipartimento di Fisica\\
Via della Ricerca Scientifica, I-00133 Roma, Italy
}

\makeatletter
\renewcommand{\@email}[1]{#1}
\makeatother

\emailAdd{\{\href{mailto:agnese.bissi@physics.uu.se}{\tt agnese.bissi},
                     \href{mailto:andrea.manenti@physics.uu.se}{\tt andrea.manenti}\}\texttt{\textcolor{blue}{@physics.uu.se}};\\
                     \{\href{mailto:fucito@roma2.infn.it}{\tt fucito},
                     \href{mailto:morales@roma2.infn.it}{\tt morales},
                     \href{mailto:savelli@roma2.infn.it}{\tt savelli}\}\texttt{\textcolor{blue}{@roma2.infn.it}}
}

\abstract{The calculation of physical quantities in certain quantum field theories such as those of the Argyres-Douglas type is notoriously hard, due to the lack of a Lagrangian description. Here we tackle this problem following two alternative approaches. On the one hand, we use localization on the four-sphere to compute two-correlators and OPE coefficients in Argyres-Douglas superconformal theories.  
On the other hand, we use the conformal bootstrap machinery to put stringent bounds on such coefficients, only relying on the knowledge of central charge and conformal dimension of the operators. We compare the results obtained with these two methods and find good agreement for all rank-one cases and for the rank-two Argyres-Douglas theories $(A_1,A_4)$ and $(A_1,A_5)$, in the moduli space of pure $SU(5)$ and $SU(6)$ super Yang-Mills. We also apply our results from localization to obtain bounds on the dimensions of the lightest neutral unprotected operators of the CFTs. 
}

\preprint{\hfill UUITP-65/21}

\begin{document}

\maketitle

\section{Introduction and summary}

This paper deals with $\mathcal{N}=2$ four-dimensional superconformal field theories (SCFT) and with the computation of the OPE coefficients of the chiral primary operators parametrizing the Coulomb branch (CB) of their moduli space. Research in the last two decades has enormously increased the landscape of such theories, which features points lacking marginal deformations, hence constituting isolated non-Lagrangian theories with an intrinsically strong interaction. The archetypes of such theories are the Argyres-Douglas (AD) and Minahan-Nemeschansky (MN) theories \cite{Argyres:1995jj,Argyres:1995xn,Eguchi:1996vu,Eguchi:1996ds,Minahan:1996cj}. More recently \cite{Argyres:2015ffa,Argyres:2015gha,Argyres:2016xmc,Argyres:2016xua} this group of theories has been further enlarged in the attempt to compile a complete list of $\mathcal{N}=2$ SCFT classifying their moduli space of vacua, characterized by their Coulomb branches, Higgs branches and possibly mixed branches. The main properties of such theories have been derived either with field theory methods thanks to their connection to gauge theories through dualities and RG-flows \cite{Argyres:2010py,Argyres:2007tq,Maruyoshi:2016aim,Agarwal:2016pjo}, with geometric techniques via class ${\cal S}$ constructions and engineering in string theory \cite{Gaiotto:2009we,Bonelli:2011aa,Xie:2012hs,Cecotti:2012jx,Cecotti:2013lda,Chacaltana:2014nya,Wang:2015mra,Chacaltana:2016shw,Giacomelli:2017ckh} or by studying the superconformal index~\cite{Buican:2015ina,Cordova:2015nma,Agarwal:2018zqi,Xie:2021omd,Song:2021dhu,Buican:2021elx}.

It is well known that the dynamics of superconformal theories is strongly constrained. Indeed, superconformal symmetry is powerful enough to determine all correlators of half-BPS operators of the theory out of the two- and three-point functions of superconformal primaries. 
 Denoting by $\{ {\cal O}_a \}$ a basis of primary operators, a SCFT is therefore specified by the conformal dimensions $\Delta_a$, the metric $g_{ab}$ and the OPE coefficients $\lambda_{abc}$, respectively defined as
\be
G_{ab}(x)  =\langle \mathcal{O}_a(x)\mathcal{O}_b(0)\rangle={g_{ab} \over |x|^{\Delta_a+\Delta_b}}\,,\qquad\qquad{\cal O}_a (x) {\cal O}_b(0) =\sum_{c}  { \lambda_{abc} {\cal O}_c(0) \over |x|^{\Delta_a+\Delta_b-\Delta_c} } +\cdots\,,
\ee
where the dots stand for subleading terms in the limit $x\to 0$. A particularly interesting class of operators are the scalar \emph{chiral} primaries $\{ {\cal O}_i \}$. This set of operators form a commutative ring ${\cal O}_i (x) {\cal O}_j(0)\sim {\cal O}_{i} (0){\cal O}_{j} (0)+\ldots$, which makes the OPE coefficients  either zero or one. The chiral/anti-chiral two-point functions $G_{ij}  =\langle \mathcal{O}_i(x)\bar{\mathcal{O}}_{j}(0)\rangle$, instead, are non-trivial. Alternatively,
in an orthonormal basis where $\hat {g}_{ij}= \delta_{ij}$, the SCFT data is encoded in the OPE coefficients $\hat{\lambda}_{ijk}$ which can be expressed in terms of the metric coefficients $G_{ij}$. The OPE coefficients $\hat \lambda_{ijk}$ will be the focus of this paper.\footnote{OPE coefficients 
will always be written in the orthonormal basis, so the hat will be omitted from now on.}
 We will compute them using two different approaches: the conformal bootstrap and localization.

  The starting point of the conformal bootstrap is the writing of the four-point function as an infinite sum of conformal blocks associated to the exchange of operators of arbitrary weight and spin, with coefficients given by quadratic combinations of the unknown OPE coefficients.
  Crossing relations are then used to bound the region of allowed values for the OPE coefficients after a specific numerical optimization procedure. This approach has been pioneered in \cite{Rattazzi:2008pe} and detailed in the supersymmetric $\mathcal{N}=2$ case in \cite{Beem:2014zpa,Cornagliotto:2017snu,Gimenez-Grau:2020jrx}. See \cite{Simmons-Duffin:2016gjk, Poland:2018epd} for recent reviews.

 The second approach we use is based on localization techniques, introduced in \cite{Nekrasov:2002qd,Flume:2002az,Bruzzo:2002xf} for gauge theories on $\mathbb{R}^4$, and later in  \cite{Pestun:2007rz} for $S^4$. Two-point correlators of chiral/anti-chiral primary operators have been computed using localization on the four-sphere, with operator insertions at the North and South poles \cite{Baggio:2014ioa,Baggio:2014sna,Baggio:2015vxa,Baggio:2016skg,Gerchkovitz:2016gxx,Rodriguez-Gomez:2016ijh,Rodriguez-Gomez:2016cem,Billo:2017glv,Billo:2019job,Beccaria:2020hgy,Beccaria:2021hvt,Billo:2021rdb}. Results at weak coupling were shown to reproduce those obtained via Feynman-diagram techniques. Instanton corrections for $SU(2)$ with $N_f=4$ massless flavours were shown to be in complete agreement with those obtained by the conformal bootstrap method \cite{Beem:2014zpa}.


In this paper, we are interested in two-point correlators of those SCFT's of the AD type, lying in the CB moduli space  of $\mathcal{N}=2$ gauge theories with matter, at points where mutually non-local degrees of freedom become simultaneously massless. Consequently these theories lack a Lagrangian description.\footnote{Some of these theories, although they have no $\cN=2$ conformal manifold, may be reached from $\cN=1$ Lagrangian theories in the UV by an RG flow \cite{Maruyoshi:2016tqk}.}

 The work is inspired by the observation \cite{Russo:2014nka,Russo:2019ipg} that, for specific choices of the masses,
   the partition function on $S^4$ of $SU(N)$ gauge theories with fundamental matter is dominated at large radius by saddle points, some of which are precisely the AD points, suggesting that the large-radius physics is described by a superconformal field theory. Here we exploit this observation
     to extract the two-point correlators of the CB operators and, consequently, their OPE coefficients for interacting non-Lagrangian conformal field theories, from the large-radius limit of gauge theories with fundamental matter.
  
  Following \cite{Billo:2017glv,Billo:2019job,Gerchkovitz:2016gxx}, we derive a localization formula for the two-point correlators on $S^4$, and cast the result in the form
  \be
 G_{ij} (x) =\langle \mathcal{O}_i(x)\bar{\mathcal{O}}_{j}(0)\rangle = C_{ij} -C_{im} C^{mn} C_{nj}\,,
\label{aqqbar20}
\ee
  where  $C_{ij}$  is the matrix-model two-point function
     \be
C_{ij}={1\over {Z}_{S^4}}
 \int d {a} \,  O_i({a},q)  \bar{O}_j({a},\bar q)  \big| Z_{\mathbb{R}^4} ({a},\epsilon_1,\epsilon_2,q)\big|^2\,,
\label{correlatorsS40}
\ee
and $x=2\pi R$, with $R$ the radius of the four sphere.
The integral is performed over the imaginary axes for ${a}=\{ a_1,\ldots,a_{N-1} \}$. $Z_{\mathbb{R}^4}$ and $O_i$ are respectively the partition function and the one-point function on $\mathbb{R}^4$  in an $\Omega$-background where $\epsilon_1 =\epsilon_2=R^{-1}$, whereas ${Z}_{S^4}$ is the partition function on the four-sphere, $q=\Lambda^{2N-N_f}/4$, $N_f$ is the number of flavours  and $\Lambda$ is the renormalization-group invariant scale.  The sum over $m,n$ in (\ref{aqqbar20}) runs over all operators with dimensions lower than $\Delta_i$ and gets rid of the operator mixing 
via Gram-Schmidt orthogonalization. 

In the large-radius limit, the integral (\ref{correlatorsS40}), for specific choices of the masses, is dominated by a saddle point corresponding to an AD conformal theory in the moduli space of the gauge theory, and $G_{ij}$ will therefore give rise to the two-point correlator of that conformal theory. 
 
     There are two main difficulties in evaluating the saddle-point integral. First, the saddle point is located at strong coupling, so one has to rely on non-perturbative formulae for the prepotential, where the entire tower of instanton contributions has been resummed. Second, the correlators receive contributions from higher-order terms in the radius expansion, codified by the generalized prepotentials ${\cal F}_{g\geq2}$, which survive the limit ${a}\to {a}_*$, $R\to \infty$ with $({a}-{a}_*)R$ finite, where $a_*$ is the saddle-point value of $a$. 
  In the case of $SU(2)$ gauge theories with matter, and the AD theories describing their strong coupling regimes, exact formulae for the SW periods are known and the contributions of the SW prepotential ${\cal F}_0$ and its first gravitational correction ${\cal F}_1$ can be evaluated explicitly. 
For higher-rank AD theories, explicit formulae for the SW periods are, to our knowledge, not known. Here we compute the SW periods for some rank-two AD theories describing the strong coupling regimes of pure $SU(5)$ and $SU(6)$ gauges theories  (see \cite{Masuda:1996np} for studies of the SW periods in the underlying SQCD theories).  The SW periods are explicitly evaluated and summed up to generalized hypergeometric functions.   
  
   While there is a priori no reason to expect that the higher gravitational terms encoded in the functions ${\cal F}_{g>1}$ are suppressed, we 
  find remarkable evidence that they give mild contributions to the OPE coefficients: Indeed, 
  we show that, in almost all cases we analyzed, results obtained including only the contributions of ${\cal F}_0$ and ${\cal F}_1$ are inside the bootstrap windows.  Similar conclusions were reached in \cite{Grassi:2019txd}, where the results for the OPE coefficients of the rank-one AD theories have been extrapolated from matrix-model formulae derived in the large-R-charge limit. 
 We show  similar agreement for the rank-two AD theories, known as $(A_1,A_4)$ and $(A_1,A_5)$, which lie in the moduli space of pure $SU(5)$ and $SU(6)$ super Yang-Mills. Thanks to the relatively low values of the conformal dimensions and the central charges, we are able to compute, using the bootstrap machinery, stringent bounds for the OPE coefficients in these theories, providing remarkable tests of the localization formula. 
    
  Finally, we observe that the localization formula for the OPE coefficients makes reference only to  intrinsic data of the SCFT, and can therefore be  extended to other non-Lagrangian SCFT's, not necessarily connected to gauge theories, like the MN ones. Unfortunately, the conformal dimensions in these theories are not so small, and the windows obtained from the conformal bootstrap are too wide to qualify as a valid test.
  
%
%
This is the plan of the paper: In Section \ref{Sec:Localization} we discuss localization and compute the OPE coefficients. 
In Section \ref{Sec:Bootstrap} we discuss the results obtained with the bootstrap. In Section \ref{Sec:LargeCharge} we verify that our formulae for computing the OPE coefficients are compatible with those valid at large R-charge available in the literature. Finally in Section \ref{Sec:Conclusions} we draw our conclusions and discuss some open questions. Technical details are deferred to the appendices.

\section{OPE coefficients from localization}\label{Sec:Localization}
This section is divided into five subsections. In \ref{Sec:LocalizationTwoPoint} we introduce the localization formula for two-point correlators on $S^4$, and discuss the large-radius limit in \ref{Sec:LargeRadius} and the integration measure in \ref{Sec:Measure}. In \ref{Sec:SU2computation} and \ref{Sec:SU56computation} we compute the OPE coefficients for the AD theories lying in the CB moduli space of $SU(2)$ SQCD and $SU(5),SU(6)$ pure super Yang-Mills respectively.

\subsection{Localization formula}\label{Sec:LocalizationTwoPoint}
\subsubsection{One-point functions on $\mathbb{R}^4$}

The partition function on $\R^4$ for a supersymmetric $\mathcal{N}=2$ $SU(N)$ gauge theory with $N_f$ flavours in an $\Omega$ background with parameters $\epsilon_1,\epsilon_2$, can be written  as the product of a tree level, a one-loop, and an instanton contribution as \cite{Nekrasov:2002qd}
\be
Z_{\mathbb{R}^4} (a,\epsilon_1,\epsilon_2,q)=Z_{\rm tree} (a,\epsilon_1,\epsilon_2,q)Z_{\rm one-loop} (a,\epsilon_1,\epsilon_2)
Z_{\rm inst} (a,\epsilon_1,\epsilon_2,q)\,,
\label{partitionR4}
\ee
where \cite{Alday:2009aq,Billo:2013fi,Fucito:2013fba}
\bea
Z_{\rm tree} (a,q) &=& q^{ \sum_u a_u^2 \over 2\epsilon_1 \epsilon_2  } \nn\\
Z_{\rm one-looop} (a) &=&   {  \prod_{u=1}^N \prod_{f=1}^{N_f}  \Gamma_2\left(a_{u}-m_i +{\epsilon\over 2}\right) \over \prod_{u < v}^N \Gamma_2(a_{uv})\Gamma_2(a_{uv}+\epsilon)  }   \nn\\
Z_{\mathbb{R}^4, \rm inst} (a,q) &=&  \sum_{Y} q^{|Y|}  Z_Y= \sum_{Y} q^{|Y|}   { \prod_{u=1}^N \prod_{f=1}^{N_f}   z_{Y_u,0} \left(a_{u}-m_i +{\epsilon\over 2}\right) \over \prod_{u , v=1}^N
z_{Y_u,Y_v} (a_{uv})  } \label{znek}
\eea
with $\epsilon=\epsilon_1+\epsilon_2$, $ q = e^{2\pi {\rm i} \tau} \mu^{2N-N_f}$ the instanton-counting parameter ($\mu$ being the renormalization reference scale) and
\be
\tau={\theta\over 2\pi  } +{4\pi {\rm i} \over g^2}\,.
\label{taucoupling}
\ee
The sum
over  $Y=\{ Y_u\}_{u=1,\ldots,N} $ runs  over the $N$-tuplets of Young tableaux with a total number $|Y|$ of boxes
  and
\bea
z_{Y_u,Y_v} &=& \prod_{(i,j)\in Y_u} \left[ x+\epsilon_1 (i-k_{vj}) -\epsilon_2(j-1-\tilde{k}_{ui} ) \right] \nn\\
&& \times \prod_{(i,j)\in Y_v} \left[ x-\epsilon_1 (i-1-k_{uj}) -\epsilon_2(j-\tilde{k}_{vi} ) \right]\,,
\label{zryy}
\eea
where $k_{uj}$ and $\tilde k_{ui}$ denote the lengths of the $j^{\rm th}$ row and $i^{\rm th}$ column of $Y_u$.

We are interested in correlators involving chiral primary operators made out of the scalar field $\varphi$ in the vector multiplet. We write
 \be
 {\cal O}_{J_1,J_2\ldots J_n} (x)=  {\rm tr\,}  \varphi^{J_1}(x) \,  {\rm tr\,} \varphi^{J_2}(x)\,  \ldots  {\rm tr\,} \varphi^{J_n}(x)\,.
 \label{chiral2}
 \ee
 Sometimes we will find it convenient to use a collective index $i$ (which refers to the total R-charge) to denote the operators in (\ref{chiral2}) as ${\cal O}_i(x)$ and their dimension as $\Delta_i$. The one-point function of such operators on ${\mathbb{R}^4}$ is given by
\be
 O_{i}(a,q) =\langle \,{\cal O}_{J_1,J_2\ldots J_n}(x)\rangle = \sum_{Y} q^{|Y|} Z_Y
 O _{J_1,Y}O _{J_2,Y}\ldots O _{J_n,Y}\,,
 \label{corrR4}
\ee
where
\be
O _{J,Y} =
 \sum_{s=1}^{|Y|}    \left[   (\chi_s{+}\epsilon_1)^J {+}(\chi_s{+}\epsilon_2)^J  {-}\chi_s^J{-}(\chi_s {+}\epsilon)^J    \right]\,,
\label{varphifix}
\ee
and
\be
\chi_s=\chi_{u,(i,j)\in Y_u}=a_u+(i-1)\epsilon_1+(j-1)\epsilon_2\,.
\ee

\subsubsection{Two-point correlators on $S^4$}

The partition function on $S^4$ of a supersymmetric $\mathcal{N}=2$ $SU(N)$ gauge theory
can be written as \cite{Pestun:2007rz}\footnote{Notice that here we have reabsorbed
the Vandermonde determinant $\prod _{u<v}^{N}a_{uv}^2$ into the modulus square of the one-loop partition function (\ref{znek})
using the identity $\Gamma(x+\epsilon_1)\Gamma(x+\epsilon_2)=x \Gamma(x)\Gamma(x+\epsilon)$   }
\begin{equation}
{Z}_{S^4}=\int d {a} \,  \big| Z_{\mathbb{R}^4} (  {a},\epsilon_1,\epsilon_2,q)\big|^2\,,
 \label{ws444}
\end{equation}
where ${a}=\{a_u\in\C \}_{u=1,\ldots,N}$, with $\sum_u a_u=0$, and the integral is performed over the imaginary axes.
$Z_{\mathbb{R}^4}$ is the partition function on $\R^4$  in an $\Omega$ background with  parameters
\be
\epsilon_1 \epsilon_2 ={1\over R^2} \, ,\qquad     b=\sqrt{\epsilon_1\over  \epsilon_2} \,,
\ee
with $R$ the radius of the sphere and $b$ the squashing parameter that here will always be set to one. 

    We now introduce the two-point matrix-model integral
\be
C_{ij}={1\over {Z}_{S^4}}
 \int d {a} \,  O_i({a},q)  \bar{O}_j({a},\bar q)  \big| Z_{\mathbb{R}^4} ({a},\epsilon_1,\epsilon_2,q)\big|^2\,,
\label{correlatorsS4}
\ee
where $O_i({a},q)$ are the one-point functions of the chiral primary operators on $\mathbb{R}^4$, defined in \eqref{corrR4}. 

Following \cite{Gerchkovitz:2016gxx,Billo:2017glv,Billo:2019job}, the two-point correlators $G_{ij}$ on the sphere can be obtained as
\be
 G_{ij}= \langle {\cal O}_i(2\pi R)\bar{\cal O}_j(0)\rangle  = C_{ij} -C_{im} C^{mn} C_{nj}\,,
\label{aqqbar2}
\ee
where the second term ``subtracts'' the mixing contributions between operators of different dimensions. Indeed, the sum over $m,n$ runs over all the operators with dimensions lower than $\Delta_{i}$
 and $C^{mn}$ is the inverse of the mixing matrix restricted to the space of such operators. It follows
 from (\ref{aqqbar2})  that $G_{in}=0$, i.e.~there is no mixing between operators of different dimensions, as expected. 

 Let us consider the OPE coefficients. Since we are dealing with chiral operators, there are no contractions between them,  and therefore
 \be
 {\cal O}_i (x) {\cal O}_j(0) =  {\cal O}_{i}(0)  {\cal O}_{j}(0)    +\ldots\,,
 \ee
 i.e.~the only non-trivial OPE coefficients are $\hat\lambda_{i,j,k_{ij} }=1$ with $k_{ij}$ the index label for 
 the operator ${\cal O}_{i} {\cal O}_{j}$. One can\ introduce the canonically-normalized operators $\hat{\cal O}_i$
 \be
{\cal O}_i (x)= \sqrt{G_{ii}} \, {\cal \hat O}_i(x)\,.
\label{rescalingO}
\ee
 In the new basis $\hat{G}_{ij}\sim\delta_{ij}/R^{2\Delta_i}$ and
\be
\hat \lambda_{i,j,k_{ij}}=\sqrt{\frac{G_{k_{ij} , k_{ij} }}{G_{ii} G_{jj}}}\,. 
\label{opecoeff}
\ee
 For the first few non-zero two-point correlators one finds
\bea
 G_{11} &=&   C_{11}{-}C_{01} C_{10} \label{gs} \nn\\ \\ 
 G_{22} &=& C_{22}{-}\frac{C_{01} C_{12} C_{20}{+}C_{02} C_{10}
   C_{21}{-}C_{02} C_{11} C_{20}{-} C_{12} C_{21}}{C_{01} C_{10}{-} C_{11}}\,,\nn
 \label{OPEcoeff1}
 \eea
  Alternatively one can write \cite{Gerchkovitz:2016gxx}
 \be
 G_{nn} ={{\rm det}_{i,j\leq n}  C_{ij}  \over {\rm det}_{i,j\leq n-1}  C_{ij} }\,.
\label{OPEcoeff2}
\ee
Using these formulae one can evaluate the two-point correlators and the OPE coefficients for any SCFT with a known prepotential.  In the
case of the $\mathcal{N}=2$ SCFT with $SU(2)$ gauge group and four fundamental massless flavours, the computation of $\lambda_{112}$ 
has been performed in \cite{Beem:2014zpa} numerically, keeping into account the first few instantonic contributions, and matched against the result obtained via the conformal bootstrap. Note that formula \eqref{OPEcoeff2} cannot be used for rank higher than one, because of the mixing of operators of the same dimension.\footnote{In the large-$N$ limit of gauge theories, the mixing of multi-trace operators can be neglected, and a formula like \eqref{OPEcoeff2} can still be used \cite{Beccaria:2020hgy,Beccaria:2021hvt,Billo:2021rdb}.} Formula \eqref{aqqbar2}, instead, can be used for any rank, leading in general to a non-diagonal matrix on the space of operators of the same dimension.

\subsection{The flat-space limit}\label{Sec:LargeRadius}

As we hinted in the Introduction, an important class of $\mathcal{N}=2$ SCFT \cite{Argyres:1995xn,Argyres:1995jj,Minahan:1996cj} live in the strong coupling regime of an asymptotically-free gauge theory. They are isolated and do not admit a Lagrangian formulation. Can localization be used in these cases? The answer is yes.
The crucial observation was put forward in \cite{Russo:2014nka,Russo:2019ipg}, where the partition function $Z_{S^4}$ of the $SU(N)$ gauge theories with fundamental matter was studied in the limit where the radius of the four-sphere is large. In this limit 
 the gauge prepotential ${\cal F}$ can be written as
 \be
 {\cal F}( a,R) \approx   \sum_{g=0}^\infty   {\cal F}_g (a) \, R^{-2g} \,,
    \label{rf0}
\ee 
 where $ {\cal F}_0 (a)$ is the SW prepotential characterizing the low energy  dynamics of the gauge theory in flat space and ${\cal F}_g$ account for the gravitational corrections arising from the $\Omega$-curvature of the spacetime. 
The integral  (\ref{ws444})  is dominated by a saddle
point at the extremum ${a}= {a}_*$ of the SW prepotential ${\cal F}_0(a)$, i.e.~where
\be
a^s_{D}(a_*)=-{1\over 2\pi {\rm i}} {\partial {\cal F}_0 \over \partial a_s} \Big|_{a=a_*}=0\,, \qquad \quad s=1,\ldots,N-1\,,
\ee
leading to $Z_{S^4} \approx \big| e^{ \Lambda^2 R^2 f_* } \big|^2$ with $ {\cal F}_0(a_*)=\Lambda^2 \, f_*$ and $f_*$ a complex number. More precisely,
based on dimensional analysis, the expansion for large $\Lambda R$ of the prepotential can be written as\footnote{We have taken all masses to be of the order of $\Lambda$, and they will eventually be tuned to specific values corresponding to the AD points.}
\be
R^2\, {\cal F} (a,\Lambda, R) \approx (\Lambda R)^2 f_* +F[ (a-a_*) R ]  \,, 
\ee
 where we discarded terms  suppressed in the limit  $\Lambda R\to\infty$.  Plugging this into (\ref{correlatorsS4}) one finds 
\be
C_{ij}\approx   
{1\over Z_{S^4}  } \int d {a} \, O_i({a}) \, O_j({a})   e^{ F[ (a-a_*) R ] +{\rm h.c.} } \,,
\ee
with
\be
 Z_{S^4} \approx  \int d{a} \, e^{ F[ (a-a_*) R ] +{\rm h.c.} } \,.
\ee
 It is important to observe that the physics near a generic saddle point is better described in terms of the dual periods $a^s_{D}$. However, if the saddle point corresponds to an AD conformal point, either $a$ or $a_D$ can be used, because they are related by
\be\label{tauaDa}
a^s_{D} \approx  \tau^{s s'} \left( a-a_{*}\right)_{s'}\,,
\ee 
with $\tau^{s s'}$ a symmetric matrix with positive-definite imaginary part, whose entries depend homogeneously on $a_s$ with degree zero.

\subsection{The u-plane integral}\label{Sec:Measure}

The SW periods $a_s(u)$ provide a relation between the variables $a=\{a_s \}_{s=1,\ldots N-1} $ and the curve parameters $u=\{u_n\}_{n=2,\ldots,N}$ representing the gauge-invariant coordinates of the CB moduli space. 
One can use these relations to write the partition function as an integral over the u-plane. The SW prepotential ${\cal F}_0(u)$ can be obtained from 
$a_D(u)$ upon u-integration
\be
{\partial {\cal F}_0(u) \over \partial u_n} =  -2\pi {\rm i} \, a_D^s(u) {\partial a_s(u)\over \partial u_n} \,.
\ee
 The first gravitational correction ${\cal F}_1$ can also be written in terms of the SW periods and the discriminant $\Delta({u})$ of the SW geometry  \cite{Nakajima:2003uh,Witten:1995gf,Moore:1997pc,Manschot:2019pog,Shapere:2008zf}
  \be
{\cal F}_1(u)   = -  \log\left[  \Lambda^{\frac{N(N-1)}{2}}  {\rm det} \left( {\partial {a_s} \over \partial {u_n}  } \right)\right]^{1 \over 2}  +\left( b^2+{1\over b^2} \right) \log\left[  {\Delta(u) \over \Lambda^{2N(2N-1)} }\right]^{1\over 24}\!\!.
\label{F1g}
\ee
The above formula can be checked perturbatively in $q$ using the Nekrasov partition function reviewed in Appendix \ref{SWSU(2)gauge}, or verified using modular invariance of the path integral.
  Plugging  (\ref{F1g}) into (\ref{correlatorsS4}) and setting the squashing parameter $b=1$ one finds
   \be
C_{ij} ={1\over {Z}_{S^4}}
 \int d{u} \, {\cal O}_i({u}) \,{\cal O}_j({u}) \,   \left|  \Delta({u})^{1\over 6}  e^{ 2 R^2  {\cal F}_0 ({u})} \right|  \,  +\ldots \,,
\label{correlatorsS43}
\ee
and  
\be
  {Z}_{S^4} =
 \int d {u}\,   \left|  \Delta({u})^{1\over 6}  e^{ 2 R^2  {\cal F}_0 ({u})} \right|  +\ldots \,,
\label{ZS43}
\ee
where  the dots are the contribution of higher gravitational terms ($g\geq2$).\footnote{The inclusion of ${\cal F}_1$ is crucial in order to reconstruct a modular covariant u-integration measure \cite{Shapere:2008zf}.} These contributions are in general not known in a closed form.
In this paper we will compute the correlators and the OPE coefficients including only the contributions of ${\cal F}_0$ and ${\cal F}_1$.
Surprisingly, the results already fall, in most of the cases, inside the narrow windows determined by the conformal bootstrap techniques, showing small discrepancies only for few operators of low dimensions.

\subsection[$SU(2)$ gauge theory with fundamentals]{$\boldsymbol{SU(2)}$ gauge theory with fundamentals}\label{Sec:SU2computation}

In this section we apply the method we previously discussed to the computation of the OPE coefficients  for the $SU(2)$ gauge theory with $N_f=1,2,3$ fundamental flavors (setting equal masses for all the flavors). The SW geometry of these theories is given by
\be
 w(x)^2=\prod_{i=1}^4 (x-e_i) =P(x)^2  -\Lambda^{4-N_f} \, (x-m)^{N_f} \,,
 \ee
with 
 \be
  P(x)=
\left\{
\begin{array}{ll}
  x^2-u \qquad& N_f=1\,,     \\
   x^2-u +{\Lambda^2\over 4}  \qquad& N_f=2 \,,    \\
    x^2-u +{\Lambda\over 4}  ( x- 3  m) \qquad& N_f=3   \,.  \\
\end{array}
\right.
\ee
The SW prepotential ${\cal F}_0$  can be obtained from the SW periods via
\be
{\cal F}_0(u) = - 2\pi {\rm i}\int^u du' a_{D} (u') {\partial a(u')\over \partial u'}\,,  \label{prepo}
\ee
with $a(u)$ and $a_D(u)$ the SW periods, whose explicit expressions are given in Appendix \ref{SWSU(2)gauge}.

We are interested in the physics near the AD conformal point, obtained by requiring that three branch points collide, i.e.
  \be
  w^2(x)=(w^{2})'(x)=(w^{2})''(x)=0\,,
  \ee
  that can be solved for $x$, $u$, and $m$.

  \begin{table}
\centering
$
\setstretch{1.3}
\begin{array}{|c|cccccc|}
\hline
N_f  & d & c & x_*  & u_* &  m_*  &  \alpha  \\
\hline
 1  & \ft{6}{5} & \ft{11}{30}&  \ft{\Lambda}{2} &  \ft{3\Lambda^2}{4}  &  -\ft{3\Lambda}{4} & 2  \\
  2  & \ft{4}{3} & \ft12 & {\Lambda\over 2}    & {\Lambda^2\over 2}  & {\Lambda\over 2}  & 3  \\
   3  & \ft{3}{2} & \ft23  &  {\Lambda\over 8} &   {5\Lambda^2\over 64}  & - {\Lambda\over 8}  & 4 \\
   \hline
\end{array}
\setstretch{1}
$
\caption{\label{tablead}Summary of the data identifying the AD theories in the CB of $SU(2)$ SQCD. $c$ denotes the central charge, $d$ the CB-operator conformal dimension, and $\alpha$ the power at which the CB coordinate appears in the discriminant.}
\end{table} 
Expanding the discriminant, given in Eq.~\eqref{ddelta}, and the prepotential in Eq.~(\ref{prepo}) around the AD saddle point one finds
  \bea
\Delta(u) & = & c_\Delta\,  \Lambda^{12 }\,\left({ u-u_* \over \Lambda^2} \right)^\alpha \,,\nn\\
    {\cal F}_0(u) &=&    \Lambda^2 \, \left[ f_*   - {c^2_{\cal F} \over 2}  \, \left({ u-u_* \over \Lambda^2} \right)^{2\over d}  \right] \,, \label{uf0}
\eea
    with $c_\Delta$ and $c^2_{\cal F}$ some real dimensionless constants, whose value is irrelevant for our purposes,\footnote{The minus sign in front of $c^2_{\cal F}$, though, is crucial in order for the integrals in \eqref{correlatorsS43} to converge. But, as can be seen by combining \eqref{uf0} with \eqref{tauaDa}, this sign is guaranteed by the positive-definiteness of the gauge coupling ${\rm Im}\, \tau$.} and $d, \alpha$ rational numbers characterizing the AD theory. The results for $N_f=1,2,3$ are displayed in Table
    \ref{tablead}. The dynamics of the gauge theory near these points is described by the rank-$1$ AD conformal field theories known as $\mathcal{H}_0$, $\mathcal{H}_1$, and $\mathcal{H}_2$ respectively. 
 In particular, one finds
   \be
   \alpha=N_f+1=12\lsp\frac{d-1}{d} \,.
   \ee
   
 Now let us consider the two-point correlators in these theories. 
 The  chiral ring is generated by a single operator  that can be taken to be
 \be
{\cal O}(x) =\Lambda^{d-2}\left( \ft12 {\rm tr} \varphi^2(x) - u_*\right) =\tilde u \,,
\ee
with  the shift chosen such that $\langle {\cal O} \rangle =0$ at the SCFT point $u=u_*$, and the normalization chosen such that all the dependence on the scale $\Lambda$ drops out in two point functions $C_{ij}$.
The remaining operators in the chiral ring are obtained as powers ${\cal O}_i(x)={\cal O}(x)^i$ of the chiral ring generator. The one-point function factorizes as
\be
O_i(u)=\langle {\cal O} (x)^i \rangle =   \tilde{u}^i\,. \label{oi}
\ee
  Plugging (\ref{uf0}) and (\ref{oi}) into (\ref{correlatorsS43}) one finds\footnote{Here we use
  \be
  \int_0^\infty e^{-c x^\beta}  x^\gamma=\beta^{-1} c^{-{1+\gamma\over \beta} } \Gamma\left( { 1+ \gamma \over \beta}  \right)\nn\,.
  \ee
    }
 \bea
C_{ij}  & \approx &  {  \int_0^\infty d\tilde{u} \,    e^{- R^2 c^2_{\cal F}\,   \tilde{u}^{\frac{2}{d}}  }  \,  \tilde{u}^{i+j+{\alpha\over 6}} \over
 \int_0^\infty d\tilde{u} \,    e^{- R^2 c^2_{\cal F} \, \hat{u}^{\frac{2}{d}}  }  \,  \tilde{u}^{\alpha \over 6}  }   ={1 \over  (R c_{\cal F})^{d(i+j) }}  { \Gamma\left(\frac{ d}{2} \left(1+i+j+\ft{\alpha}{6}  \right) \right) \over   \Gamma\left(\frac{d}{2}\left(1+\frac{\alpha}{6} \right)\right)  }\,.
\label{cij2}
\eea
 Notice that $C_{ij}\sim  R^{-{(i+j)d}}$
 as expected from conformal invariance.
 
 The two-point correlators follow then from (\ref{aqqbar2}) or (\ref{OPEcoeff2}), and the OPE coefficients from (\ref{opecoeff}).  The results are displayed in Table \ref{tSW}. 
 
   Interestingly our results depend only on two numbers, $d$ and $\alpha$, characterizing the dimension of the CB operator and 
 the degree of the discriminant of the curve. In particular, the latter is related to the central charge $c$ of the theory. These two numbers are codified in the SW curves and can be easily computed for non-lagrangian theories like the MN theories with flavour symmetry $E_6$, $E_7$, and $E_8$.
 
 This data will be compared in Section \ref{Sec:Bootstrap} against the results obtained using the bootstrap approach.
\begin{table}
\centering
$
\setstretch{1.2}
\begin{array}{|c|cccccc|}
\hline
{\rm SCFT} & \mathcal{H}_0 & \mathcal{H}_1&\mathcal{H}_2 &E_6 &E_7 & E_8\\
\hline
 {\rm SW} & y^2=x^3{+} {u} & y^2=x^3{+} {u} x & y^2=x^3{+} {u}^2& y^2=x^3{+} {u}^4& y^2=x^3{+} {u}^3 x& y^2=x^3{+} {u}^5\\
 \hline
 d  & \frac{6}{5} & \frac{4}{3} & \frac{3}{2} & 3 & 4 & 6 \\
 c & \frac{11}{30} & \frac{1}{2} & \frac{2}{3} & \ft{13}{6} & \ft{19}{6} & \ft{31}{6} \\
  \alpha &2& 3 & 4 &  8 &9& 10\\
\hline
\lambda^2_{u\lsp u\lsp u^2} &  2.09823 & 2.24125 & 2.42063 & 4.51365 & 6.75467 & 15.1158 \\
\lambda^2_{u\lsp u^2\lsp u^3}  & 3.30002 & 3.67408 & 4.17529 & 12.0469 & 24.011 & 95.3327 \\
\lambda^2_{u^2\lsp u^2\lsp u^4} &7.20621 & 8.62414 & 10.7157 & 67.008 & 222.212 & 2443.47 \\
 \hline
\end{array}
\setstretch{1}
$
\caption{\label{tSW}Some OPE coefficients for the rank-$1$ AD theories $\mathcal{H}_0,\mathcal{H}_1,\mathcal{H}_2$, and for the rank-$1$ MN theories $E_6,E_7,E_8$. We have removed the $\tilde{}$ from $u$ in order not to clutter the notation.}
\end{table}

\subsection[AD conformal points in pure \texorpdfstring{$SU(N)$}{SU(N)} gauge theories]{AD conformal points in pure $\boldsymbol{SU(N)}$ gauge theories}

\label{Sec:SU56computation}

The SW geometry describing the low energy dynamics of pure $SU(N)$ gauge theories  can be written as
 \be
 w(x)^2 =P_N(x)^2-\Lambda^{2N}\,,
 \ee
where 
\be
P_N=x^N-u_2 x^{N-2}-u_3 x^{N-3} \ldots -u_N
\ee
and $u_n$ the CB parameters. These theories contain an AD conformal point \cite{Eguchi:1996ds}, usually called $(A_1,A_N)$, at 
\be
 u_2=u_3=\ldots u_{N-1}= 0 \, , \qquad u_N=\Lambda^N \,,
\ee
 which is obtained by setting
 \be
 x = \tilde x  \nu\,, \qquad \qquad w = \tilde w  \, {\rm i} \sqrt{2} \,(\Lambda\nu)^{N/2} \,, \qquad \qquad u_n =\tilde u_n \, \nu^{n} +\delta_{n,N} \Lambda^N\,,
 \ee
 and keeping the leading order in the limit $\nu \to 0$. This leads to the AD curve
 \be
 \tilde w(x)^2 = \tilde x^N-\tilde u_2 \, \tilde x^{N-2} -\tilde u_3 \tilde x^{N-3}  \ldots -\tilde u_N\,, \label{adsw}
  \ee
   and the SW differential  
\be
\lambda = {\tilde w d \tilde x} \,.
\ee
  The conformal dimensions of the parameters $\tilde u_n$ entering the curve  is determined by requiring that the SW differential has dimension one
$d(\lambda)=1$, leading to
\be
d(\tilde x) ={2\over N+2}\,, \qquad \qquad  d(\tilde w) ={N\over N+2}\,, \qquad \qquad  d(\tilde u_n) ={2 n\over N+2}  \,.\label{puregaugedims}
\ee
 These parameters are interpreted as Coulomb branch parameters, masses and couplings depending on whether dimensions are bigger, equal or lower than 1 respectively.  In particular Coulomb branch parameters $\tilde u_n$ correspond to the choices
 \be
 N\geq n > {N\over 2}+1\,. \label{rint}
 \ee
 There are $r=\left[ \ft{N-1}{2} \right] $ of such operators, with $r$ often referred to as the rank of the CFT and $[A]$ denoting the integral part of $A$.  Finally the central charge of the CFT  is given by
 \cite{Shapere:2008zf}
 \be\label{puregaugec}
 c={d(\Delta) \over 12} +{r\over 6} =  {N(N-1)\over 6(N+2) } + \ft16 \left[ \frac{N-1}{2} \right]\,.
 \ee 
    
\begin{table}
	\setstretch{1.25}
	\centering
	$
	\begin{array}{|c|c|c|c|c|c|c|}
	\hline
	N & d(\tilde u_n) & {\rm rank}   & {\rm curve}&  {\rm CFT} & \Delta & c \\
	\hline
	3& {4\over 5}  ,   {\bf {6\over 5}} &1  & \tilde w^2=\tilde x^3- v  & \mathcal{H}_0\equiv(A_1,A_2)  & -3^3\, v^2 & \ft{11}{ 30}  \\ 
	4 &  {2\over 3} , 1   , {\bf {4\over 3}}&1 & \tilde w^2=\tilde x^4 -v  & \mathcal{H}_1\equiv(A_1,A_3) & -4^4\, \, v^3 &  \ft12\\
	5 & {4\over 7} , {6\over 7} , {\bf {8\over 7}}  , {\bf {10\over 7}}  &2  & \tilde w^2=\tilde x^5-u\, x-\, v & (A_1,A_4)& -4^4 \, u^5+5^5\, v^4  &\ft{17}{21} \\
	6 & ~~ {1\over 2} , {3\over 4} , 1, {\bf {5\over 4}} , {\bf {3\over 2}}  ~~&2& ~~ \tilde w^2=\tilde x^6-u\, x-v~~ & (A_1,A_5)&  5^5 \, u^6+6^6 \, v^5 & \ft{23}{24} \\
	\hline
	\end{array}\
	\setstretch{1}
	$
	\caption{\label{ttSW}CB conformal dimensions and SW curves for the $(A_1,A_N)$ theories with $N=3,\ldots,6$. We set masses and couplings to zero.}
\end{table}

In Table \ref{ttSW} we list the conformal dimensions  $d(\tilde u_n)$ of the parameters entering the SW curve for the theories of rank up to two, indicating in boldface those operators  
spanning the Coulomb branch. We display also the curves (and their determinant) obtained by setting to zero masses and couplings, and renaming the CB operators as $u$, $v$. For $N=3,4$, one finds the rank-one CFTs  $\mathcal{H}_0$, $\mathcal{H}_1$ studied in the previous section.  For $N=5,6$ one finds the rank-two AD CFTs known as $(A_1,A_4)$ and $(A_1,A_5)$. 

 In a rank-two theory, one can consider correlators involving a single Coulomb branch operator or mixed ones. In the former case, the correlator can be naively computed, by integrating over the one-dimensional slice defined by deformations from the conformal point involving only the CB operator we are interested in. The two-point correlators $G_{nn}$ are then given (\ref{OPEcoeff2}) with $C_{ij}$ given by the rank-one formula
 (\ref{cij2}). The resulting OPE coefficients, say for correlators involving only $u$-insertions, depend then only on the dimension $d(\tilde u)$ and the exponent $\alpha_u $ defining the vanishing rate of the discriminant $\Delta \sim u^{\alpha_u }$. The results obtained from this one-dimensional truncation are displayed in Table \ref{tablesu560}. 
While we do not know how to physically justify such a truncation, we will see that it nevertheless gives results in good agreement with the bootstrap.

   \begin{table}
  \setstretch{1.2}
  	\centering
  	$
  	\begin{array}{|c|cc|cc|c|c|c|}
  	\hline
  	&~ d(u)~&~ \alpha_u~ & ~d(v)~ &~ \alpha_v~ & ~c ~ &  \lambda^2_{u\, u\, u^2}     & \lambda^2_{v\,v\,v^2}     \\
  	\hline
  	(A_1,A_4)   & \ft{8}{7} & 5 &  \ft{10}{7} & 4  & \ft{17}{21} & 1.981  & 2.327 \\
  	(A_1,A_5)     & \ft{5}{4} & 6 & \ft{3}{2} & 5 & \ft{23}{24}  & 2.077    & 2.378\\
  	\hline
  	\end{array}
  	$
   \setstretch{1}
  	\caption{\label{tablesu560} OPE coefficients for the rank-$2$ AD theories of type $(A_1,A_N)$ obtained by truncating the Coulomb branch 
	integral to a single operator slice.}
  \end{table}
   
 In the next two sections we will derive the OPE coefficients (including mixed ones) from a full-fledged two-dimensional integration over the CB moduli space.

\subsubsection{$(A_1,A_5)$ Argyres-Douglas theory}
 
  After setting the masses and couplings to zero, the low energy dynamics of $(A_1,A_5)$ at a generic point in the Coulomb branch is described by the curve\footnote{This analysis generalizes the results of \cite{Eguchi:1996vu} in this case.}
  \be
  \tilde w^2 = \tilde x^6 -  u \tilde x - v \,.
  \ee
 The SW periods are defined by
 \be
 a_s =\oint_{\alpha_s}  \lambda \,,\quad \qquad   a_D^s =\oint_{\beta_s} \lambda\,, \label{swperiods}  
 \ee
with 
\be
\lambda =d\tilde x \sqrt{ \tilde x^6 -  u \tilde x - v }\,,
\ee
$(\alpha_s,\beta_s)$ is a basis of cycles with intersection matrix $\alpha_s  \cap \beta_{s'}=\delta_{s s'}$ and $s,s'=1,2$. The integrals 
(\ref{swperiods}) can be explicitly performed in the limit of small $u$. In this limit the branch points are located around the sixth roots
  \be
  e_i =v^{1\over 6} \, w^i 
  \ee 
  \  with $w = e^{\pi {\rm i}  \over 3}$. The SW differential can be expanded as
  \be
 \lambda =d\tilde x \sum_{n=0}^\infty  {\Gamma(\ft32)\, (-u \tilde x)^n \over n! \Gamma(\ft32-n) }  \,  (\tilde x^6 - v)^{ {1\over 2}-n }
  \ee
 leading to the integrals
  \be
  \Pi_i =\oint_{\gamma_i}  \lambda=2 \int_{e_i}^{e_{i+1}}  \lambda = {\rm i} \, w^i\, V^{2\over 3}   A( w^i \kappa)  \,,   \ee
   where 
   \be
   A(\kappa)=\sqrt{\pi }      \sum_{n=0}^\infty  
  \frac{
   \Gamma \left(\frac{n+7}{6}\right)   \kappa^n \,    \sin\left( {(n+1)\pi \over 6}  \right)  }{ (n{+}1)! \Gamma
   \left(\frac{5}{6} (2{-}n)\right)    }   \label{akappasu6}
   \ee
   and
   \be
  {\rm i} V^{2\over 3}= v^{2\over 3}  w^{{1\over 2}}  \qquad ,\qquad \kappa=u v^{-{5 \over 6}  } w^{1\over2}\,.
   \ee  
    The variables $V$ and $\kappa$ are chosen such that the $a_s$'s are purely imaginary for $V$ and $\kappa$ real and positive.
  This choice follows from \cite{Pestun:2007rz} and is dictated by the mathematical consistency of the integral over $S^4$ and the gluing of the two charts. 
     The  sum  in the right hand side of  (\ref{akappasu6}) can be carried out  and written in terms of  hypergeometric functions but the 
   explicit formula is not very illuminating, so we will omit it here.   
  The periods  can then be written as
     \be
  a_1=\Pi_{0} \,, \qquad\quad a_2=\Pi_{3}\,, \qquad\quad  a_D^1=\Pi_{1}  \,, \qquad\quad a_D^2=\Pi_{4} \,.
  \ee
 As a consistency check of our choices, we have verified that the Riemann bilinear identities are satisfied, leading to a symmetric $\tau$ matrix, with positive imaginary part.
 Furthermore we notice that the $a_s$'s are purely imaginary for $\kappa$ and $V$ real and positive.  
  
  The SW prepotential ${\cal F}_0(a)$ can be computed in terms of the SW periods. Indeed, using the fact that ${\cal F}_0(a)$ is a local homogeneous
  function of degree two in the $a$'s, one finds
 \be
 {\cal F}_0(a)=\frac{1}{2}\sum_{s=1}^2 a_s {\partial\over \partial a_s} {\cal F}_0(a) = -\pi {\rm i} \sum_{s=1}^2 a_s \, a_{D}^s \,. 
  \label{faad}
 \ee
  We introduce the function
   \be
    f(\kappa)  = - 2\, V^{-{4\over 3}} \, {\rm Re} \, {\cal F}_0 =-{\rm Re} \left\{ 2  \pi {\rm i} \,w\,  \left[  A( \kappa)   A( w\kappa)  +  A( -\kappa)   A( -w\kappa) \right] \right\}\,.
  \ee
  One can check that $f(\kappa)$ is a positive, monotonously increasing function for $\kappa\geq0$. 
      The matrix-model type integral (\ref{correlatorsS43}) reduces to
 \bea
 C_{u^{m} v^n}&=& \frac{1}{Z_{S^4}}\int d \kappa \,dV  \left|  6^6{-} 5^5 \kappa^6 \right|^{1\over 6} 
\, e^{ -V^{4\over 3} \, f(\kappa)}  \, V^{ {5\over 3} +{5i\over 6} +n} \kappa^m \nn\\
&=&  \frac{1}{Z_{S^4}} \int d \kappa   \left| 6^6 {-}5^5 \kappa^6 \right|^{1\over 6} 
\,  \Gamma\left(  2{+}\ft{5m}{8}{+}\ft{3n}{4}\right)  \left[ f(\kappa)\right]^{ -2-\ft{5m}{8}-\ft{3n}{4}} \,  \kappa^m\,. 
\eea
 The last integral is performed numerically.  The results are displayed in Table \ref{tablesu56}.

   \begin{table}
  \setstretch{1.2}
  	\centering
  	$
  	\begin{array}{|c|c|c|c|c|c|c|}
  	\hline
  	&~ [u]~ & ~[v]~ & ~c ~ &  \lambda^2_{u\,\llsp u\,\llsp u^2}  &\lambda^2_{u\,\llsp v\,\llsp u v}  & \lambda^2_{v\,\llsp v\,\llsp v^2}     \\
  	\hline
  	(A_1,A_4)   & \ft{8}{7} & \ft{10}{7} & \ft{17}{21} &1.878  & 1.043  & 2.231 \\
  	(A_1,A_5)     & \ft{5}{4} & \ft{3}{2} & \ft{23}{24}  & 1.929  & 1.039  & 2.203 \\
  	\hline
  	\end{array}
  	$
   \setstretch{1}
  	\caption{\label{tablesu56} Some OPE coefficients for the rank-$2$ AD theories of type $(A_1,A_N)$.}
  \end{table}

\subsubsection{$(A_1,A_4)$ Argyres-Douglas theory}
 
  After setting masses and couplings to zero, the low energy dynamic of $(A_1,A_4)$ at a generic point in the CB is described by the curve
  \be
  \tilde w^2 = \tilde x^5 -  u \tilde x - v \,.
  \ee
 The SW differential is now
\be
\lambda =d\tilde x \sqrt{ \tilde x^5 -  u \tilde x - v }\,.
\ee
 Again we computed the integrals as an expansion for $u$ small. In this limit the branch points are located around the fifth roots
  \be
  e_i =v^{1\over 5} \, w^i 
  \ee 
  \  with $w=e^{2\pi {\rm i}  \over 5}$. The SW differential can be expanded as
  \be
 \lambda =d\tilde x \sum_{n=0}^\infty  {\Gamma(\ft32)\, (-u \tilde x)^n \over n! \Gamma(\ft32-n) }  \,  (\tilde x^5 - v)^{ {1\over 2}-n }
  \ee
 leading to the integrals
  \be
  \Pi_i =\oint_{\gamma_i}  \lambda=2 \int_{e_i}^{e_{i+1}}  \lambda = {\rm i} \, w^i\, V^{7\over 10}   A( w^i \kappa)\,,     \ee
   where 
   \be
   A(\kappa)=\sqrt{\pi }      \sum_{n=0}^\infty  
  \frac{
   \Gamma \left(\frac{n+6}{5}\right)  \kappa^n \,    \sin\left( {(n+1)\pi \over 5}  \right)  }{ (n{+}1)! \Gamma
   \left(\frac{1}{10} (17{-}8n)\right)    }   \label{akappasu5}
   \ee
   and
   \be
  {\rm i} V^{7\over 10}= v^{7\over 10}  w^{{1\over 2}}  \, ,\quad\qquad \kappa=u v^{-{4 \over 5}  } w^{1\over2}\,.
   \ee   
  The periods  are now given by
     \be
  a_1=\Pi_{0} \,, \qquad\quad a_2=\Pi_{4}+\Pi_1\,, \qquad\quad  a_D^1=\Pi_{1}  \,, \qquad\quad a_D^2=-\Pi_{3}  \,.
  \ee
  The SW prepotential ${\cal F}_0(a)$ is again computed from (\ref{faad}) leading to
   \be
    f(\kappa)  ={ -} 2\, V^{-{7\over 5}} \, {\rm Re} \, {\cal F}_0 =-{\rm Re} \left\{ 2  \pi {\rm i} \,\left[ w\,   A( \kappa)   A( w\kappa)  {-} w^2\,   A( w^3 \kappa)   A(w^4 \kappa) {-} w^4\,    A( w^3 \kappa)   A(w \kappa) \right] \right\}\,.
  \ee
 Once again $f(\kappa)$ is a positive, monotonously increasing function  for $\kappa\geq0$. 
      The matrix-model integral (\ref{correlatorsS43}) reduces to
 \bea
 C_{u^{m} v^n}&=&\frac{1}{Z_{S^4}} \int d \kappa \,dV  \left|  5^5{+} 4^4 \kappa^5 \right|^{1\over 6} 
\, e^{ -V^{7\over 5} \, f(\kappa)}  \, V^{ {22\over 15} +{4m\over 5} +n} \kappa^m \nn\\
&=&  \frac{1}{Z_{S^4}} \int d \kappa   \left| 5^5 {+}4^4 \kappa^5 \right|^{1\over 6} 
\,  \Gamma\left(  \ft{37}{21}{+}\ft{4m}{7}{+}\ft{5n}{7}\right)  \left[ f(\kappa)\right]^{ {-} {\ft{37}{21}}{-}\ft{4m}{7}{-}\ft{5n}{7}    } \,  \kappa^m\,.
\eea
 The last integral is performed numerically.  The results are displayed in Table \ref{tablesu56}.

\section{Conformal bootstrap approach}\label{Sec:Bootstrap}

In this section we would like to obtain upper and lower bounds on the OPE coefficients that have been computed in the previous section and summarised in Table \ref{tSW}. The approach that we follow is the conformal bootstrap that allows us to find bounds for the dimensions and OPE coefficients of intermediate operators by exploiting the symmetries of the specific CFT. In our setup, such consistency conditions come from superconformal symmetry, unitarity and crossing symmetry of the four-point functions. The numerical study of these conditions give bounds on the CFT data. While we defer the details to the following sections, we would like to emphasize that the bounds that we find are fully non\nobreakdash-perturbative and they do not refer to any Lagrangian description.

This section is divided into seven subsections. In Subsection~\ref{sec:multiplets} we show the multiplets exchanged in the OPE of two Coulomb branch operators. In Subsection~\ref{sec:bootstrapEqns} we write down the crossing equations and then we discuss their numerical implementation in Subsection~\ref{sec:numerical}. Subsections~\ref{sec:OPE_bounds} and~\ref{sec:gap_bounds} explain the setup for obtaining bounds on the OPE coefficients and operator dimensions, respectively, while Subsections~\ref{sec:res_bounds} and~\ref{sec:res_gaps} show the results we obtained.

\subsection{Superconformal multiplets exchanged}\label{sec:multiplets}
The OPE of two chiral operator has been investigated in~\cite{Fitzpatrick:2014oza, Beem:2014zpa, Lemos:2015awa, Gimenez-Grau:2020jrx}. In this section, we review their results using the notation introduced in~\cite{Cordova:2016emh}. A superconformal multiplet is fully specified by the Cartan eigenvalues of its superconformal primary. The Cartan eigenvalues consist in the conformal dimension $\Delta$, the left and right spins $j,\jb$, the $SU(2)$ Cartan $R$ and the $U(1)$ charge $r$. This information is encoded as
\eqn{
L\overbar{L}[j\lsp;\jb]^{(R;r)}_\Delta\,.
}[]
In principle one can infer any possible shortening condition by the values of $\Delta,R,r,j,\jb$. For convenience, however, shortening conditions will be indicated by replacing $L$ or $\overbar{L}$ with the symbols $A_1$, $A_2$, $B_1$ or $\bar{A}_1$, $\bar{A}_2$, $\bar{B}_1$. The meaning of these symbols has to do with the states that become null and are thus factored out of the multiplet.

The chiral and antichiral operators are the superconformal primaries of the following multiplets\footnote{In~\cite{Cordova:2016emh}'s conventions $Q$ has $\mathfrak{u}(1)$ charge $-1$ so chiral operators must have charge $\pm 2\Delta$. For simplicity, we rescale the R-charge by $-\tfrac12$ so that $r$ will denote both the conformal dimension and the charge of $\phi_r$. The multiplet notation, however, will stay consistent with its original work. In addition we emphasize that the $SU(2)$ R-charge is expressed with the Dynkin label (i.e. R-spin $\frac12$ means $R=1$.)}
\eqn{
\phi_r \in B_1\overbar{L}[0\lsp;0]_r^{(0;-2r)} \,,\qquad
\phib_r \in L\overbar{B}_1[0\lsp;0]_r^{(0;2r)} \,.
}[]
For example for a SCFT gauge theory with ${\cal N}=2$ supersymmetry, $ \phi_r$ can represent 
the primary ${\rm tr}\, \varphi^{r}(x)$ (or a multi-trace operator made of $r$ scalars) and the same for  $\phib_r $ with chiral fields replaced by anti-chiral ones.

The OPE of a chiral and an antichiral operator depends on the sign of the two charges. Assuming first that $r_1 \geq r_2 +1$ and letting $r_{12} = r_1-r_2$ one has \cite{Lemos:2015awa}
\eqn{
\phi_{r_1} \times \phib_{r_2} =  B_1\overbar{L}[0\lsp;0]_{r_{12}}^{(0;-2r_{12})} + A_t\overbar{L}[\ell\lsp;\lnsp\ell]^{(0;-2r_{12})}_{\ell+2+r_{12}} +  L\overbar{L}[\ell\lsp;\lnsp\ell]^{(0;-2r_{12})}_\Delta 
\,,
}[]
with $\Delta > \ell+2+r_{12}$ and $t = 1$ if $\ell>0$ and $t = 2$ if $\ell=0$. Since we do not have permutation symmetry, $\ell$ can assume all non-negative even and odd values. If instead $0 < r_1 - r_2 < 1$ the $B_1\overbar{L}$ multiplet is absent, while the rest of the OPE remains the same. If $r_2 > r_1$ we have the same conclusion with the conjugate multiplets while if $r_1=r_2\equiv r$ we have \cite{Fitzpatrick:2014oza, Beem:2014zpa}
\eqn{
\phi_r \times \phib_r =  A_2\overbar{A}_2[0\lsp;0]^{(0;0)}_{2} +  L\overbar{L}[\ell\lsp;\lnsp\ell]^{(0;0)}_\Delta 
\,,
}[]
with, again, $\Delta > \ell+2$. Also here $\ell$ can be even or odd.

The interesting operators that appear are the chiral operators themselves in the OPE with different R-charges and the stress tensor multiplet $A_2\overbar{A}_2[0\lsp;0]_2^{(0;0)}$. There are no higher spin analogs $A_1\overbar{A}_1[\ell\lsp;\lnsp\ell]$ as they contain higher spin currents that appear only in free theories~\cite{Maldacena:2011jn,Boulanger:2013zza,Alba:2013yda}.

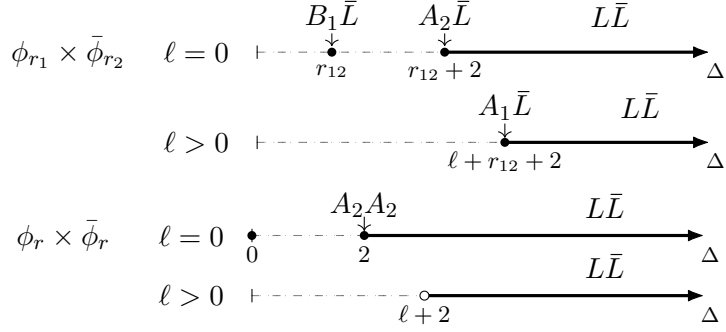
\begin{figure}
\newcommand{\phirphiminusrOPE}{\begin{tikzpicture}
\node at (-.5,0) {$\phi_{r_1} \times \phib_{r_2}$};
\node at (1.2,0) {$\ell = 0$};
\draw [thin, gray, dash pattern={on 3pt off 5pt}] (2,0) -- (7.99,0);
\node at (8.1,-.25) {\scriptsize{$\Delta$}};
\draw [{|[scale=1.1]}->, >={Latex[round, scale=1.3]}, thin, dash pattern={on 0pt off 3pt}] (2,0) -- (8,0);
\draw [very thick] (4.5,0) -- (7.88,0);
\node at (4.5,0)[circle,fill,inner sep=1.2pt]{};
\node at (3,0)[circle,fill,inner sep=1.2pt]{};
\node at (3,.5) {$B_1\overbar{L}$};
\node at (3,-.25) {\footnotesize $r_{12}$};
\node at (4.5,.5) {$A_2\overbar{L}$};
\node at (4.5,-.25) {\footnotesize $r_{12}+2$};
\draw [->] (4.5,.25)--(4.5,.08);
\draw [->] (3,.25)--(3,.08);
\node at (6.7,.5) {$L\overbar{L}$};
\begin{scope}[shift={(0,-1.2)}]
\node at (1.2,0) {$\ell > 0$};
\draw [thin, gray, dash pattern={on 3pt off 5pt}] (2,0) -- (7.99,0);
\node at (8.1,-.25) {\scriptsize{$\Delta$}};
\draw [{|[scale=1.1]}->, >={Latex[round, scale=1.3]}, thin, dash pattern={on 0pt off 3pt}] (2,0) -- (8,0);
\draw [very thick] (5.3,0) -- (7.88,0);
\node at (5.3,0)[circle,fill,inner sep=1.2pt]{};
\node at (5.3,.5) {$A_1\overbar{L}$};
\node at (5.3,-.25) {\footnotesize $\ell+r_{12}+2$};
\draw [->] (5.3,.25)--(5.3,.06);
\node at (7.1,.5) {$L\overbar{L}$};
\end{scope}
\end{tikzpicture}}
\newcommand{\phirphiminusrSameOPE}{\begin{tikzpicture}
\node at (-.5,0) {$\phi_{r} \times \phib_r$};
\node at (1.2,0) {$\ell = 0$};
\draw [thin, gray, dash pattern={on 3pt off 5pt}] (2,0) -- (7.99,0);
\node at (8.1,-.25) {\scriptsize{$\Delta$}};
\draw [{|[scale=1.1]}->, >={Latex[round, scale=1.3]}, thin, dash pattern={on 0pt off 3pt}] (2,0) -- (8,0);
\draw [very thick] (3.5,0) -- (7.88,0);
\node at (3.5,0)[circle,fill,inner sep=1.2pt]{};
\node at (2.01,-.25) {\footnotesize $0$};
\node at (2.006,0)[circle,fill,inner sep=1.2pt]{};
\node at (3.5,.4) {$A_2A_2$};
\node at (3.5,-.25) {\footnotesize $2$};
\draw [->] (3.5,.25)--(3.5,.06);
\node at (6.7,.4) {$L\overbar{L}$};
\begin{scope}[shift={(0,-.8)}]
\node at (1.2,0) {$\ell > 0$};
\draw [thin, gray, dash pattern={on 3pt off 5pt}] (2,0) -- (7.99,0);
\node at (8.1,-.25) {\scriptsize{$\Delta$}};
\draw [{|[scale=1.1]}->, >={Latex[round, scale=1.3]}, thin, dash pattern={on 0pt off 3pt}] (2,0) -- (8,0);
\draw [very thick] (4.3,0) -- (7.88,0);
\node at (4.3,0)[circle,draw=black, fill=white,inner sep=1.2pt]{};
\node at (4.3,-.25) {\footnotesize $\ell+2$};
\node at (6.7,.4) {$L\overbar{L}$};
\end{scope}
\end{tikzpicture}}
\centering
\phirphiminusrOPE
\\
\phirphiminusrSameOPE
\caption{OPE in the $\phi_{r_1} \times \phib_{r_2} $ channel depicted according to the conformal blocks $\CG^t_{\Delta,\ell}$ or $\CG^{u}_{\Delta,\ell}$ exchanged. The white dot implies that there is no operator there.}\label{fig:neutral_channel}
\end{figure}

The superconformal blocks encoding the contributions of the above multiplets, including also the whole towers of superdescendants, to a given four-point functions are all expressed in terms of a single function. This function has been denoted $\CG^{t}_{\Delta,\ell}$ for the $t$-channel correlator and $\CG^{u}_{\Delta,\ell}$ $u$-channel correlator. In Section~\ref{sec:bootstrapEqns} we give the explicit definitions. In Figure~\ref{fig:neutral_channel} we depict the ranges of $\Delta$ and $\ell$ that correspond to a given multiplet. The fact that, for example, the long multiplet half-line starts at the stress tensor point means that, from this four-point functions' point of view, the stress tensor multiplet is indistinguishable from a long multiplet at dimension very close to two. The same phenomenon will appear also elsewhere. This is important when applying numerical bootstrap techniques, and it is the reason why it is necessary to input a small gap in the long sector whenever one wants to assume something about a protected operator which sits at the unitarity bound.

\begin{figure}
\newcommand{\phirphirOPE}{\begin{tikzpicture}
\node at (-.5,0) {$\phi_{r_1} \times \phi_{r_2}$};
\node at (1.2,0) {$\ell = 0$};
\draw [thin, gray, dash pattern={on 3pt off 5pt}] (2,0) -- (7.99,0);
\node at (8.1,-.25) {\scriptsize{$\Delta$}};
\draw [{|[scale=1.1]}->, >={Latex[round, scale=1.3]}, thin, dash pattern={on 0pt off 3pt}] (2,0) -- (8,0);
\draw [very thick] (4.8,0) -- (7.88,0);
\node at (4.8,0)[circle,fill=black!40!blue!30!white,draw,inner sep=1.2pt]{};
\node at (2.8,0)[circle,fill,inner sep=1.2pt]{};
\draw [draw=black!30!red!80!white,thick,dash pattern={on 5pt off 1.8pt}, join=round, cap=round, dash phase=2pt] (2.2,-.45) rectangle ++(1.18,1.22);
\node at (2.8,.5) {$B_1\overbar{L}$};
\node at (2.8,-.25) {\footnotesize $r_1+r_2$};
\node at (4.8,.5) {$B_1\overbar{L}^{(2)}$};
\node at (4.8,-.25) {\footnotesize $r_1+r_2+2$};
\draw [->] (4.8,.25)--(4.8,.08);
\draw [->] (2.8,.25)--(2.8,.08);
\node at (6.7,.5) {$L\overbar{L}^{(4)}$};
\begin{scope}[shift={(0,-1.2)}]
\node at (1.2,0) {$\ell = 1$};
\draw [thin, gray, dash pattern={on 3pt off 5pt}] (2,0) -- (7.99,0);
\node at (8.1,-.25) {\scriptsize{$\Delta$}};
\draw [{|[scale=1.1]}->, >={Latex[round, scale=1.3]}, thin, dash pattern={on 0pt off 3pt}] (2,0) -- (8,0);
\draw [very thick] (5.2,0) -- (7.88,0);
\node at (5.2,0)[circle,fill,inner sep=1.2pt]{};
\node at (3.2,0)[circle,fill,inner sep=1.2pt]{};
\node at (3.2,.5) {$B_1\overbar{L}^{(1)}$};
\node at (3,-.25) {\footnotesize $r_1+r_2+1$};
\node at (5.2,.5) {$A_2\overbar{L}^{(3)}$};
\node at (5.2,-.25) {\footnotesize $r_1+r_2+3$};
\draw [->] (5.2,.25)--(5.2,.08);
\draw [->] (3.2,.25)--(3.2,.08);
\node at (6.7,.5) {$L\overbar{L}^{(4)}$};
\end{scope}
\begin{scope}[shift={(0,-2.4)}]
\node at (1.2,0) {$\ell > 1$};
\draw [thin, gray, dash pattern={on 3pt off 5pt}] (2,0) -- (7.99,0);
\node at (8.1,-.25) {\scriptsize{$\Delta$}};
\draw [{|[scale=1.1]}->, >={Latex[round, scale=1.3]}, thin, dash pattern={on 0pt off 3pt}] (2,0) -- (8,0);
\draw [very thick] (5.6,0) -- (7.88,0);
\node at (5.6,0)[circle,fill,inner sep=1.2pt]{};
\node at (3.6,0)[circle,fill,inner sep=1.2pt]{};
\node at (3.6,.5) {$A_t\overbar{L}^{(2)}$};
\node at (3.6,-.25) {\footnotesize $r_1+r_2+\ell$};
\node at (5.6,.5) {$A_1\overbar{L}^{(3)}$};
\node at (5.6,-.25) {\footnotesize $\quad r_1+r_2+\ell+2$};
\draw [->] (5.6,.25)--(5.6,.08);
\draw [->] (3.6,.25)--(3.6,.08);
\node at (6.9,.5) {$L\overbar{L}^{(4)}$};
\end{scope}
\end{tikzpicture}}
\centering
\phirphirOPE
\caption{OPE in the $\phi_{r_1} \times \phi_{r_2} $ channel depicted according to the conformal blocks $\CG^s_{\Delta,\ell}$ exchanged. When present, the superscript denotes (twice) the difference between the R-charge of the superconformal primary and $r_1+r_2$, which also corresponds to the level of the superdescendant that enters the OPE. The light blue dot implies that the operator is there only if a mixed branch is present. We also framed in red the the operator whose OPE coefficient we want to extremize.}\label{fig:charged_channel}
\end{figure}
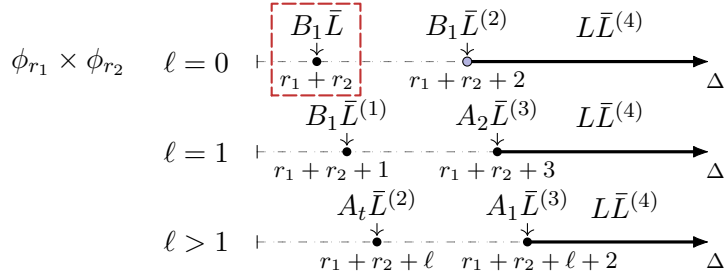

Now let us move to the OPE of two chiral operators. The list of exchanged superconformal multiplets is now much richer. In Figure~\ref{fig:charged_channel} we depict the range of spins and conformal dimensions of the blocks exchanged in this channel. The superconformal blocks reduce to usual conformal blocks $g_{\Delta,\ell}$ in this case because the selection rules allow the exchange of only one operator per multiplet. For notational consistency, however, we will denote them as $\CG^s_{\Delta,\ell}$. In the equation below all superconformal multiplets, with the exception of the first one, contribute to the OPE through a superdescendant rather than a superprimary. For example, in a long multiplet $\CO$ only $Q^4\CO$ will contribute as it is the only chiral operator in it. The OPE reads~\cite{Lemos:2015awa}
\eqna{
\phi_{r_1} \times \phi_{r_2} &=
B_1\overbar{L}[0\lsp;0]^{(0;-2(r_1+r_2))}_{r_1+r_2}+B_1\overbar{L}[0\lsp;0]^{(2;-2(r_1+r_2)+2)}_{r_1+r_2+1} + \\&\;\quad +
B_1\overbar{L}[0\lsp;\lnsp1]^{(1;-2(r_1+r_2)+1)}_{r_1+r_2+\frac12} + A_t\overbar{L}[\ell\lsp-2;\lnsp\ell]^{(0;-2(r_1+r_2)+2)}_{r_1+r_2-1+\ell} + \\&\;\quad +
A_t\overbar{L}[\ell-1\lsp;\lnsp\ell]^{(1;-2(r_1+r_2)+3)}_{r_1+r_2+\frac12+\ell} + L\overbar{L}[\ell\lsp;\lnsp\ell]_{\Delta-2}^{(0;-2(r_1+r_2)+4)}\,,
}[chargedOPE]
where $t=2$ if the first spin label is zero and $t=1$ otherwise.
The first multiplet appearing in~\eqref{chargedOPE} is the one whose OPE coefficients we want to extremize. The second one is necessarily present if there is a mixed branch in the theory. Assuming its absence requires in practice to assume a gap on the scalar long multiplets.\footnote{The numerical results are weakly affected by assuming the absence of this class of operators.}

\subsection{Bootstrap equations}\label{sec:bootstrapEqns}

In order to find the relations constraining the OPE coefficients that we are interested in, we need to study four-point correlators.
In this section we will review the crossing equations of the system of correlators involving only $\phi_r$ and $\phib_r$, which were first obtained in~\cite{Beem:2014zpa}. In appendix~\ref{app:mixed_crossing_eqns} we will also show the crossing equations of the mixed correlator system involving two pairs of chiral fields $\phi_{r_1}$, $\phi_{r_2}$ which first appeared in~\cite{Lemos:2015awa}.\footnote{In the context of superconformal field theories, mixed correlators have been studied in \cite{Lemos:2015awa,Gimenez-Grau:2020jrx,Agmon:2019imm,Bissi:2020jve}} The latter setup is needed to obtain OPE bounds on theories of higher rank or on coefficients of the type $\lambda_{u\,u^2\,u^3}^2$.

Let us start with four-point functions involving only operators $\phi_r$ and $ \phib_r$. The only nonzero ones have an equal number of chiral and antichiral fields. We will use different OPEs to define the functions $f_{s,t,u}$ as follows
\eqn{
\langle \phib_r(x_1)\phi_r(x_2)\phi_r(x_3)\phib_r(x_4)\rangle =  \frac{f_t(z,\zb)}{(x_{12}^2)^{r}(x_{34}^2)^{r}}=\,\frac{f_s(1-z,1-\zb)}{(x_{23}^2)^{r}(x_{14}^2)^{r}}=\frac{f_u\big(\frac{z}{z-1},\frac{\zb}{\zb-1}\big )}{(x_{12}^2)^{r}(x_{34}^2)^{r}}\,,
}[]
with $x_{ij}^2=|x_i-x_j|^2$ and
\eqn{
\frac{x_{12}^2\lsp x_{34}^2}{x_{13}^2\lsp x_{24}^2} = u = z\zb\,,\qquad
\frac{x_{14}^2\lsp x_{23}^2}{x_{13}^2\lsp x_{24}^2} = v = (1-z)(1-\zb)\,.
}[]
The functions $f_{s,t,u}(z,\zb)$ must satisfy these crossing constraints
\twoseqn{
((1-z)(1-\zb))^r\, f_t(z,\zb) = (z\zb)^r \, f_s(1-z,1-\zb)\,,
}[eqFirst]{
((1-z)(1-\zb))^r\, f_u(z,\zb) = (z\zb)^r \, f_u(1-z,1-\zb)\,.
}[][]
In order to write these constraints in a form amenable to numerical computations we must expand the functions $f_{s,t,u}$ in conformal blocks. Based on the analysis of section~\ref{sec:multiplets} we have\footnote{We remind the reader that $\lambda_{\CO_1\CO_2\CO_3}$ stands for the coefficient of $\CO_3$ in the OPE of $\CO_1\times\CO_2$, which means that it is computed by the three-point function $\langle\CO_1\CO_2\COb_3\rangle$.}
\threeseqn{
f_t(z,\zb) &= \CG^t_{0,0}(z,\zb)+ \frac{r^2}{6\llsp c}\CG^t_{2,0}(z,\zb) + \sum_{\ell = 0}^\infty\sum_{\Delta > \ell + 2} |\lambda_{\phi_r\phib_r\CO_{\Delta,\ell}}|^2(-1)^\ell\,\CG^t_{\Delta,\ell}(z,\zb)\,,
}[]{
f_s(z,\zb) &= |\lambda_{\phi_r\phi_r\phi_{2r}}|^2 \CG^s_{2r,0}(z,\zb) + \sum_{\substack{\ell=0\\\mathrm{even}}}^\infty \sum_{\substack{\Delta = \ell + 2r\\\Delta \geq \ell + 2r+2}} |\lambda_{\phi_r\phi_r\CO_{\Delta,\ell,2r}}|^2 \CG^s_{\Delta,\ell}(z,\zb)\,,
}[chargedExp]{
f_u(z,\zb) &= \CG^u_{0,0}(z,\zb) + \frac{r^2}{6\llsp c}\CG^u_{2,0}(z,\zb) + \sum_{\ell = 0}^\infty\sum_{\Delta > \ell + 2} |\lambda_{\phi_r\phib_r\CO_{\Delta,\ell}}|^2\,\CG^u_{\Delta,\ell}(z,\zb)\,,
}[]
where $c$ is the central charge. It is not necessary to input $c$ explicitly: it can be left arbitrary just like the other OPE coefficients. However we will fix it to a specific values on all our computations. In the first line we used $\lambda_{\phib\phi\CO} = (-1)^\ell\lambda_{\phi\phib\CO}$ and in the second we used $\lambda_{\phib\phib\COb} = \lambda_{\phi\phi\CO}^*$. We denote the operators as $\CO_{\Delta,\ell}$ if they are R-neutral, otherwise as $\CO_{\Delta,\ell,r}$, with $r$ denoting the R-charge (in units where $\phi_r$ has R-charge $r$). Notice that from this point on, the subscripts labeling $\lambda$ refer to the $r$-charge of the corresponding operators. These conventions are more suitable in this description, being also consistent with previous literature. 
 
The blocks $\CG^s,\CG^t$ and $\CG^u$ are explicit functions given by
\eqn{
\CG^I_{\Delta,\ell}(z,\zb) = \frac{z\zb}{z-\zb}\bigl(k^I_{\Delta+\ell}(z)k^I_{\Delta-\ell-2}(\zb)- z\leftrightarrow\zb\bigr)\,,\qquad I=s,t,u\,,
}[blockDef]
with
\threeseqn{
k^s_\beta(z) &= 
z^{\frac{\beta}2}\,{}_2F_1\lnsp\mleft(\frac{\beta}2,\frac{\beta}2;\beta;z\mright)\,,}[ksDef]{
k^t_\beta(z) &= 
z^{\frac{\beta}2}\,{}_2F_1\lnsp\mleft(\frac{\beta}2,\frac{\beta}2;\beta+2;z\mright)\,,}[]{
k^u_\beta(z) &= 
z^{\frac{\beta}2}\,{}_2F_1\lnsp\mleft(\frac{\beta}2,\frac{\beta+4}2;\beta+2;z\mright)\,.
}[][kDef]
With all these ingredients in place, we can finally write down the crossing equation as follows
\eqna{
&\sum_{\Delta,\ell} |\lambda_{\phi_r\phi_r\CO_{\Delta,\ell,2r}}|^2  \,\vec{V}^{\mathrm{charged}}_{\Delta,\ell}  + 
\sum_{\Delta,\ell}(\lambda_{\phi_r\phib_r\CO_{\Delta,\ell}})^2 \,\vec{V}^{\mathrm{neutral}}_{\Delta,\ell}=-\vec{V}^{\mathrm{neutral}}_{0,0} - 
\frac{r^2}{6\llsp c}\;\vec{V}^{\mathrm{neutral}}_{2,0}\,,
}[ceqSingle]
with
\eqn{
\vec{V}^{\mathrm{charged}}_{\Delta,\ell} \equiv \begin{bmatrix}
-(-1)^\ell \CF^s_{+,\Delta,\ell}  \\
(-1)^\ell \CF^s_{-,\Delta,\ell}  \\
0
\end{bmatrix}\,,\quad
\vec{V}^{\mathrm{neutral}}_{\Delta,\ell} \equiv \begin{bmatrix}
(-1)^\ell \CF^{\llsp t}_{+,\Delta,\ell}  \\
(-1)^\ell \CF^{\llsp t}_{-,\Delta,\ell}  \\
\CF^u_{-,\Delta,\ell}
\end{bmatrix}\,,
}[vvvvv]
where we introduced the (anti)symmetric combinations (which implicitly depend on $r$)
\eqn{
\CF^I_{\pm,\Delta,\ell}(z,\zb) = v^r \CG^I_{\Delta,\ell}(z,\zb)  \pm u^r \CG^I_{\Delta,\ell}(1-z,1-\zb)\,,\qquad I=s,t,u\,.
}[Ffunctions]
The crossing equation~\ceqSingle can be thought of as a vector equation in an infinite dimensional space (the space of 3-vector valued functions of $z$ and $\zb$). In the next section we will review the standard numerical methods adopted to study it.

\subsection{Numerical implementation}\label{sec:numerical}
In this section, we will give a quick review on the numerical implementation of the crossing equation~\ceqSingle. The general strategy is to find contradictions to the bootstrap equations that stem from the existence of a functional $\alpha$ satisfying some properties. Schematically, $\alpha$ should satisfy some inequalities that take the form
\eqn{
\exists\;\alpha\quad\mathrm{s.t}\quad \forall\;(\Delta,\ell) \in \mathrm{spectrum}\quad\alpha[\vec{V}_{\Delta,\ell}] \geq 0\quad \mathrm{and}\quad \alpha[\vec{V}_{0,0}^\mathrm{neutral}] = 1\,.
}[assumptions]
This would make the left hand side of~\ceqSingle non-negative and the right hand side strictly negative, hence a contradiction, meaning that the assumptions made on the spectrum were inconsistent. There are essentially two obstacles that one needs to overcome in order to translate this problem in something that can be made easily accessible by computers
\begin{enumerate}
\item Searching in a space of functionals which is infinite dimensional (i.e. the dual of a space of functions)
\item Imposing an inequality over an infinite, and in any case not fully specified, set of $\Delta$'s and $\ell$'s
\end{enumerate}
To overcome 1.\ we can restrict ourselves to a finite dimensional subspace of functionals. Empirically it turns out that the following set works particularly well\footnote{Here we show the action of $\alpha$ on a single entry of the vector $\vec{V}$. The actual $\alpha$ we are looking for is a vector of functionals, each acting on a different entry of $\vec{V}$.}
\eqn{
\alpha[F] = \sum_{n\leq m}^{n+m\leq\Lambda} \alpha_{n,m}\frac{\partial^n}{\partial z^n}\frac{\partial^m}{\partial \zb^m}F(z,\zb)\big|_{z=\zb=\frac12}\,.
}[alphaSpace]
This way, if $\Lambda$ is odd, the space of functionals has dimension
\eqn{
\mathrm{dim}(\Lambda) = \frac14(\Lambda+1)(\Lambda+3)\,.
}[]
The choice of $z=\zb=1/2$ is motivated by the fact that it is the point where both the $s$ channel OPE ($z,\zb\sim0$) and $t$ channel OPE ($z,\zb\sim1$) converge optimally. In principle one could evaluate the derivatives in slightly different points as well, but this has shown no significant improvements in the past.

To overcome 2.\ we need to work more. In principle one could discretize $\Delta$ and put a cutoff on both $\Delta$ and $\ell$ and impose positivity on all the points in this two-dimensional lattice. Despite being possibile, this is not computationally very convenient. A better approach takes advantage of the analytic structure of the conformal blocks. It is a known fact that conformal blocks and their derivatives in $z,\zb$ can be systematically approximated as a positive function times a polynomial in $\Delta$ \cite{Poland:2011ey}
\eqn{
\partial_z^n\partial_\zb^m\,{\cal G}_{\Delta,\ell}\big(\tfrac12,\tfrac12\big) = \chi(\Delta)\,p^{m,n}_\ell(\Delta)\,,\qquad
\chi(\Delta) > 0 \quad \forall\;\Delta \geq \ell+2\,.
}[eq:blockFactoriz]
The actual expression for $\chi(\Delta)$ will be shown later~\eqref{eq:derivApprox}. Because of this, we can ignore $\chi(\Delta)$ for the purpose of imposing $\alpha[g_{\Delta,\ell}] \geq 0$. Therefore, we only have to study positivity of the polynomial $p^{m,n}_\ell$ for all $\Delta \geq \ell+2$. This is a known problem in mathematics that can be mapped to imposing positive-semidefiniteness on a pair of matrices.\footnote{Very roughly, any polynomial $P(x)$ which is positive for $x>x_*$ can be written as $$P(x)=\vec{u}^T(x)\cdot M \cdot \vec{u}(x) + (x-x_*)\, \vec{u}^T(x)\cdot N \cdot \vec{u}(x)\,,$$ where $u(x)$ is a vector of independent polynomials, whose length is the degree of $P$ divided by two, and $M,N$ are positive definite matrices.}
We refer the reader to Appendix \ref{appendixrecursion} for details.

To summarize, we consider a finite subspace of functionals $\alpha$ given by~\alphaSpace and we take rational approximations of the blocks. Next we put a cutoff $\ell_{\mathrm{max}}$ on the number of spins and impose~\assumptions on all spins up to it. The rational approximation of the blocks is what allows us to impose positivity for all $\Delta$'s in one go.

\subsection{Bounds on the OPE coefficient}\label{sec:OPE_bounds}
In this specific instance of the bootstrap we want to put upper and lower bounds on the OPE coefficient of $\lambda_{\phi_r\phi_r\phi_{2r}}$. It is possible to get lower bounds only if the operator is isolated. Indeed for the theories that we are considering this is the case, as one can easily see from section~\ref{sec:multiplets}. The bootstrap problem that we want to study is
\begin{quote}\itshape
Fix $c$, then maximize (minimize) $B_{\pm} \equiv \alpha\mleft[\vec{V}^\mathrm{neutral}_{0,0} + \frac{r^2}{6c} \, \vec{V}_{2,0}^{\mathrm{neutral}}\mright]$ subject to
\begin{enumerate}
\item $\alpha[\vec{V}_{\Delta,\ell}^{\mathrm{charged}}] \geq 0\quad \forall\;\ell \geq 0\;\mathrm{even}\,,\quad \forall\;\Delta = \ell+2r\;\;\mathrm{or}\;\;\Delta \geq \ell + 2r+2$
\item $\alpha[\vec{V}_{\Delta,\ell}^{\mathrm{neutral}}] \geq 0\quad \forall\;\ell > 0\,,\quad \forall\;\Delta \geq \ell + 2$
\item $\alpha[\vec{V}_{\Delta,0}^{\mathrm{neutral}}] \geq 0\quad  \forall\;\Delta \geq 2 + \epsilon$
\item $\alpha[\vec{V}_{4,0}^{\mathrm{charged}}] = \pm 1$
\end{enumerate}
\end{quote}
The parameter $\epsilon$ can be taken to be $\sim 0.1$ and it is needed to make $V_{2,0}^{\mathrm{neutral}}$ isolated so that it is meaningful to impose a value of the central charge. 
The final result will be the two-sided bound
\eqn{
B_- \leq |\lambda_{\phi_r\phi_r\phi_{2r}}|^2 \leq B_+\,.
}[]
There will be such a bound for every value of $r$ and $c$, so we can compute it for any rank-one $\CN=2$ SCFT or any other higher-rank SCFT if we focus only on a single Coulomb branch operator. Bounds involving mixed correlators are also possible but we do not show the list of bootstrap assumptions for brevity.

\subsection{Bound on the dimension of the lightest neutral unprotected operator}\label{sec:gap_bounds}
If we know the OPE coefficient of the chiral scalar of dimension $2r$ we can put it in the right hand side of the bootstrap equation. The assumptions that we need are as follows

\begin{quote}\itshape
Fix the value of $\lambda_{\phi_r\phi_r\phi_{2r}}$ and of $c$. Then find $\alpha$ such that
\begin{enumerate}
\item $\alpha\mleft[\vec{V}^\mathrm{neutral}_{0,0} + |\lambda_{\phi_r\phi_r\phi_{2r}}|^2\lsp \vec{V}^{\mathrm{charged}}_{2r,0} + \frac{r^2}{6c} \, \vec{V}_{2,0}^{\mathrm{neutral}}\mright]=1$
\item $\alpha[\vec{V}_{\Delta,\ell}^{\mathrm{neutral}}] \geq 0\quad \forall\;\ell > 0\,,\quad \forall\;\Delta \geq \ell + 2$
\item $\alpha[\vec{V}_{\Delta,\ell}^{\mathrm{charged}}] \geq 0\quad \forall\;\ell \geq 2\;\mathrm{even}\,,\quad \forall\;\Delta = \ell+2r\;\;\mathrm{or}\;\;\Delta \geq \ell + 2r+2$
\item $\alpha[\vec{V}_{2r+2,0}^{\mathrm{charged}}] \geq 0$
\item $\alpha[\vec{V}_{\Delta,0}^{\mathrm{neutral}}] \geq 0\quad \forall\;\Delta \geq \Delta_\mathrm{gap} > 2$
\end{enumerate}
\end{quote}

The gap $\Delta_{\mathrm{gap}}$ can be taken to be inside an interval $[\Delta_\mathrm{min},\Delta_\mathrm{max}]$ and we can run a binary search by checking $\Delta_\mathrm{gap} = \frac12(\Delta_\mathrm{min}+\Delta_\mathrm{max})$ and updating the upper or lower limit according to whether an $\alpha$ satisfying the above assumptions is found or not.

\subsection{Results for the OPE coefficient}\label{sec:res_bounds}

In Tables~\ref{tab:summary1} and \ref{tab:summary2}  we report our results for the OPE bounds obtained via the numerical bootstrap. The first and last line use the single correlator setup described in Section~\ref{sec:bootstrapEqns}. The second line uses the mixed correlator setup instead, whose crossing vectors are shown in appendix~\ref{app:mixed_crossing_eqns}.

In order to obtain these bounds we set up a bootstrap problem as explained in section~\ref{sec:OPE_bounds}. We then use a newly developed framework, \texttt{sailboot},\footnote{Available at \href{https://gitlab.com/maneandrea/sailboot}{\texttt{gitlab.com/maneandrea/sailboot}}.} to translate crossing equations and assumptions into numerical vectors that can be fed to \texttt{sdpb}.\footnote{Available at \href{https://github.com/davidsd/sdpb}{\texttt{github.com/davidsd/sdpb}}~\cite{Landry:2019qug}.}

The upper and lower bounds for the simplest OPE coefficient in the $\mathcal{H}_0$ theory, namely $\lambda_{\phi_r\,\phi_r\phi^2_{r}}$ at $r=\frac65$, were previously obtained in~\cite{Cornagliotto:2017snu} while bounds for the same OPE coefficient but in the $\mathcal{H}_1$ and $\mathcal{H}_2$ theories first appeared in~\cite{Gimenez-Grau:2020jrx}. We also did the same computation for Minahan-Nemeschansky theories but, unfortunately, the bounds were either extremely weak or non-existent. This is generally expected when one considers correlators of operators with large external dimensions. A possible cause of this is that when the external dimensions are large there is a big gap between the unitarity bound and the generalized free theory spectrum, leading to many ``fake'' approximate solutions to crossing that spoil the numerics.\footnote{We thank Alessandro Vichi for discussions about this point.}

\begin{table}[h]
\centering

\setstretch{1.05}
\begin{tabular}{|r|ccc|}
\hline
theory & $\mathcal{H}_0$           & $\mathcal{H}_1$            & $\mathcal{H}_2$          
  \\
$(r,c)$ & $(\frac65,\frac{11}{30})$ & $(\frac43,\frac12)$ & $(\frac32,\frac23)$ 
\\[2pt]
\hline
\multirow{2}{*}{
$\lambda^2_{u\,u\,u^2}$} & 2.167 & 2.359 & 2.698 
\\
                                    & 2.142 & 2.215 & 2.298 
                                     \\ \hline
\multirow{2}{*}{
$\lambda^2_{u\,u^2\,u^3}$} & 3.637 & 4.445 &\\
                                      & 3.192 & 3.217 &\\ \hline
\end{tabular}
\caption{Bootstrap bounds on OPE coefficients of some rank-one theories obtained with the single correlator setup at $\Lambda=43$ ($\lambda^2_{u\,u\,u^2}$) and the mixed correlator setup at $\Lambda=19$ ($\lambda^2_{u\,u^2\,u^3}$). \label{tab:summary1}}

\setstretch{1}
\end{table}

Next we considered some theories of rank two. In Table~\ref{tablesu56} we show the results for some theories of interest. The OPE coefficients that we considered are either $\lambda_{u\,v\,uv}^2$, where $u\equiv \phi_{r_1}$ and $v\equiv \phi_{r_2}$ are the two Coulomb branch operators, or $\lambda_{\phi_i\,\phi_i\,\phi_i^2}^2$ with $\phi_i$ either $u$ or $v$. For the former one needs to use the mixed correlator setup, described in appendix~\ref{app:mixed_crossing_eqns}, whereas for the latter the single correlator setup is sufficient. It should be noted that when studying $\lambda_{\phi_i\,\phi_i\,\phi_i^2}^2$, the information that the theory is of rank two is not fed into the bootstrap equations, if not indirectly through its central charge. Some more results for the OPE coefficient bounds of heavier operators in the pure $SU(5)$ theory can be found in Table~\ref{tab:rank2more}.

\begin{table}[h]
\centering

\setstretch{1.05}
\begin{tabular}{|r|cc|}
\hline
theory & $(A_1,A_4)$ & $(A_1, A_5)$   \\
$(r_u,r_v,c)$ & $(\frac87, \frac{10}7, \frac{17}{21})$ & $(\frac54, \frac32, \frac{23}{24})$  \\[2pt]
\hline

\multirow{2}{*}{
$\lambda^2_{u\,u\,u^2}$}  & 2.102 & 2.231 \\
                               & 2.024 & 2.055 \\ \hline

\multirow{2}{*}{
$\lambda^2_{u\,v\,uv}$} & 1.125 & 1.233 \\
                                 & 0.981 & 0.960 \\ \hline

\multirow{2}{*}{
$\lambda^2_{v\,v\,v^2}$}  & 2.533 & 2.709 \\
							  & 2.181 & 2.195 \\\hline
									 
\end{tabular}
\caption{Bootstrap bounds on OPE coefficients of some rank-two theories obtained with the single correlator setup at $\Lambda=41$ ($\lambda^2_{u\,u\,u^2}$ and $\lambda^2_{v\,v\,v^2}$) and the mixed correlator setup at $\Lambda=19$ ($\lambda^2_{u\,v\,uv}$).  \label{tab:summary2}}

\setstretch{1}
\end{table}

\begin{table}[h]

\centering
\setstretch{1.05}
\begin{tabular}{|r|ccc|}
\hline&&&\\[-12pt]
coefficient & $\lambda^2_{u\,u^2\,u^3}$  & $\lambda^2_{u^2\,u^2\,u^4}$ & $\lambda^2_{v^2\,v^2\,v^4}$ \\[3pt]
\hline

\multirow{2}{*}{
bounds}          & 3.512  & 12.42 & 142.6 \\
                 & 0.890  & 2.871 & 0.855 \\

\hline
									 
\end{tabular}
\caption{Bootstrap bounds on OPE other coefficients of the rank-two theory $(A_1,A_4)$ with $c=17/21$, $r_1=8/7$ and $r_2=10/7$ obtained with the single correlator setup at $\Lambda=41$ ($\lambda^2_{u^2\,u^2\,u^4}$ and $\lambda^2_{v^2\,v^2\,v^4}$) and the mixed correlator setup at $\Lambda=19$ ($\lambda^2_{u\,u^2\,u^3}$).\label{tab:rank2more}}

\setstretch{1}
\end{table}

\begin{figure}
\centering
\includegraphicsWlabel{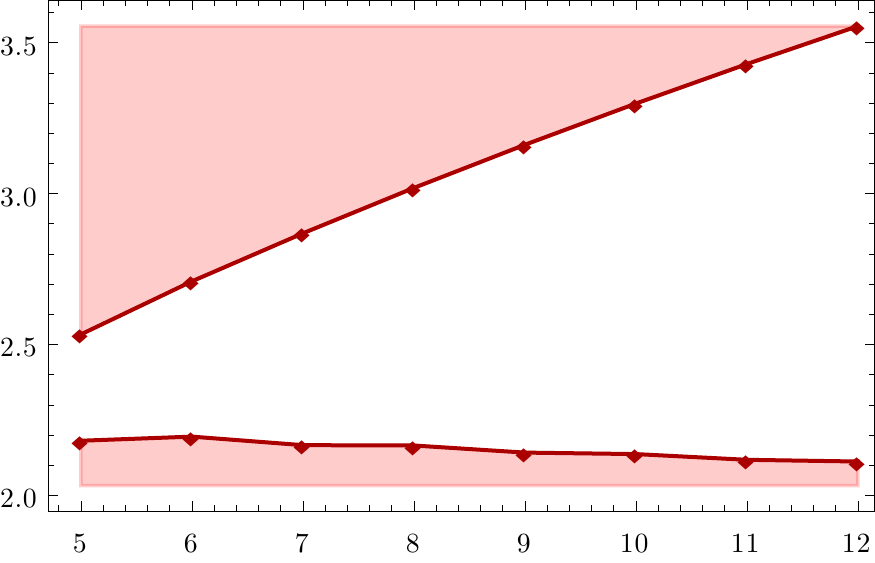}{$\lambda_{r\,r\,2r}^2$}{$N$}
\caption{Upper and lower bounds for the OPE coefficients $\lambda_{r\,r\,2r}^2$ in the pure $SU(N)$ theories where $r$ is the conformal dimension of the heaviest Coulomb branch generator. The shaded area is disallowed.}\label{fig:pureSUN}
\end{figure}

The pure~$SU(5)$ theory belongs to a family of theories of incresing rank all obtained by going to a special locus of a pure gauge theory with gauge group $SU(N)$~\cite{Martone:2021ixp,Eguchi:1996vu,Eguchi:1996ds}. The central charges and Coulomb branch dimensions are summarized in equations~\eqref{puregaugec} and~\eqref{puregaugedims}, respectively.
In Figure~\ref{fig:pureSUN} we plot upper and lower bounds for the OPE coefficients $\lambda_{u\,u\,u^2}^2$ of the theories above for $N$ ranging from $5$~to~$12$ where $r$ is the dimension of the heaviest Coulomb branch generator, namely $r=2N/(N+2)$. 

\begin{figure}
\centering
\subfloat[{$(A_1,A_4)$}]{\includegraphics[scale=1]{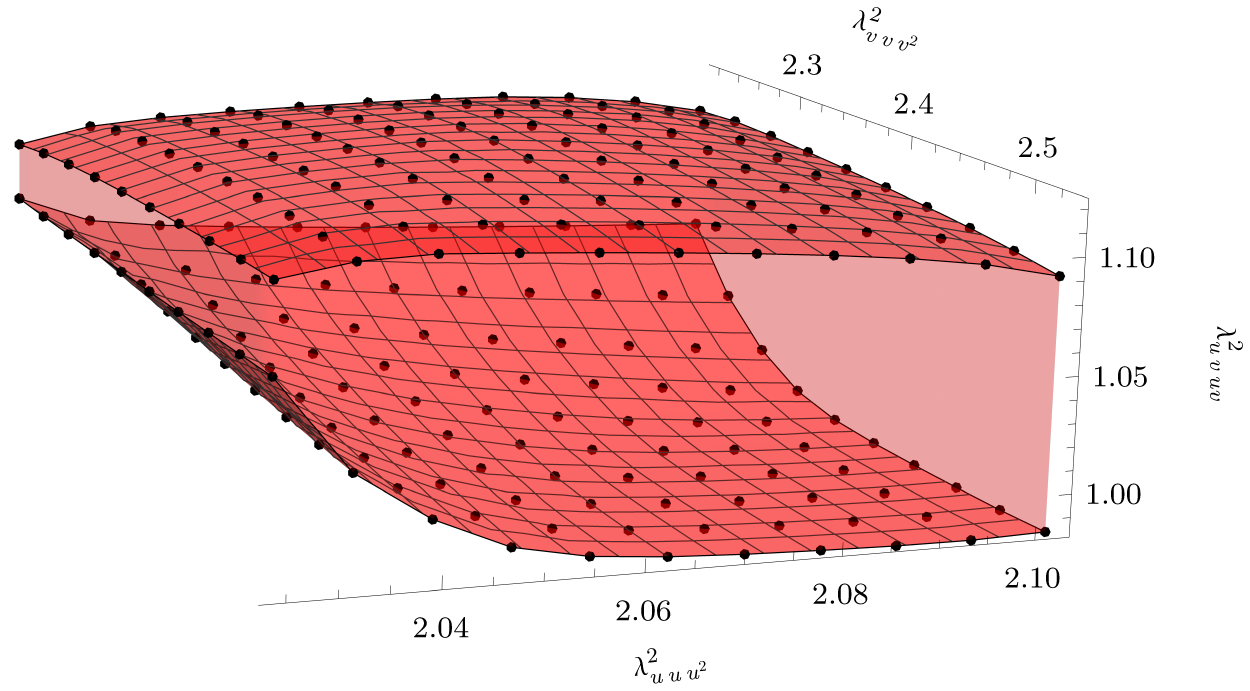}}\\
\subfloat[{$(A_1,A_5)$}]{\includegraphics[scale=1]{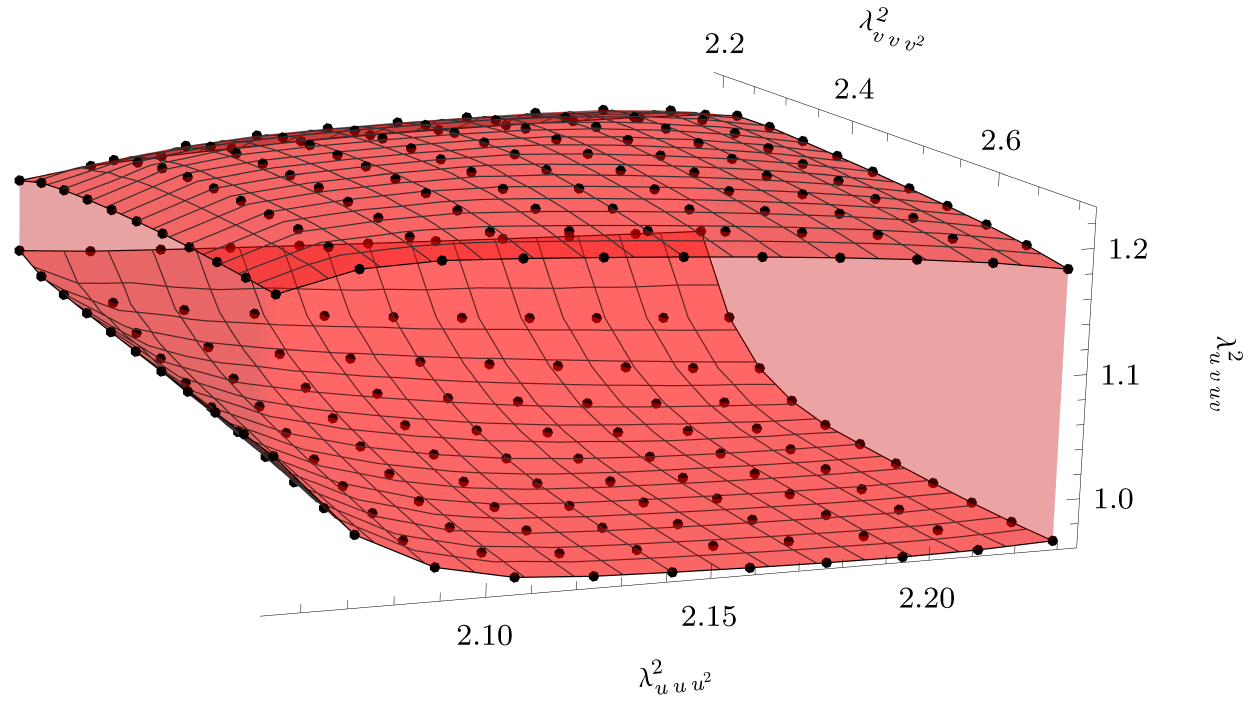}}
\caption{Upper and lower bounds on $\lambda_{u\,v\,uv}$ as a function of the values of $\lambda^2_{u\,u\,u^2}$ and $\lambda^2_{v\,v\,v^2}$. The base of the plot is bounded by the values found in the single correlator bootstrap in Table~\ref{tab:summary2}.}
\label{fig:shell}
\end{figure}

For rank-two, we can carve out the cube of OPE bounds by scanning over the values of $\lambda^2_{u\,u\,u^2}$, $\lambda^2_{v\,v\,v^2}$ and obtaining for each point an upper and a lower bound on $\lambda^2_{u\,v\,uv}$. This can be done by putting the vectors $\vec{V}^{2r_i}_{2r_i,0}$ of~\eqref{mixedVecDef} to the right of the crossing equations, just like the first point in Subsection~\ref{sec:gap_bounds}. The results for the $(A_1,A_n)$ theories for $n=4,5$ can be found in Figure~\ref{fig:shell}.

\subsection{Results for the lightest scalar neutral unprotected operator}\label{sec:res_gaps}

\begin{figure}
\centering
\subfloat[Theory $\mathcal{H}_0$, $r=\tfrac65$]{\includegraphicsWlabel[scale=1.03]{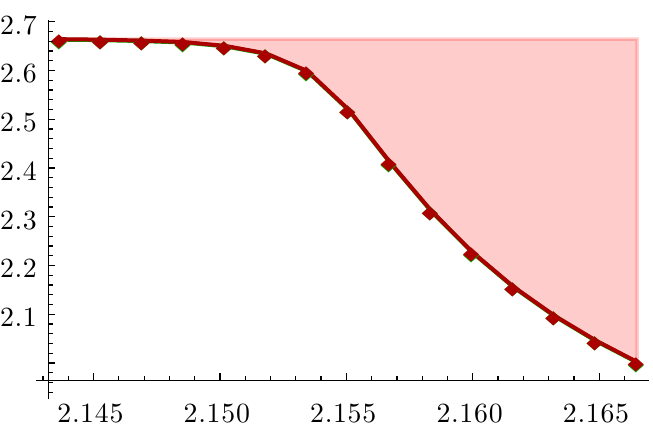}{$\Delta_{\mathrm{gap}}$}{$\lambda^2_{u\,u\,u^2}$}}\,
\subfloat[Theory $\mathcal{H}_1$, $r=\tfrac43$]{\includegraphicsWlabel[scale=1.03]{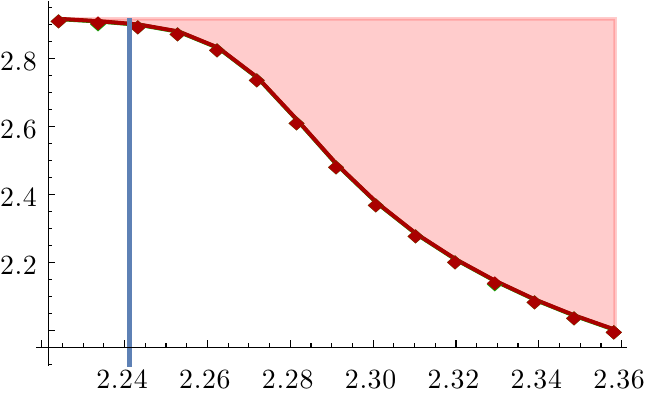}{$\Delta_{\mathrm{gap}}$}{$\lambda^2_{u\,u\,u^2}$}} \\
\subfloat[Theory $\mathcal{H}_2$, $r=\tfrac32$]{\includegraphicsWlabel[scale=1.03]{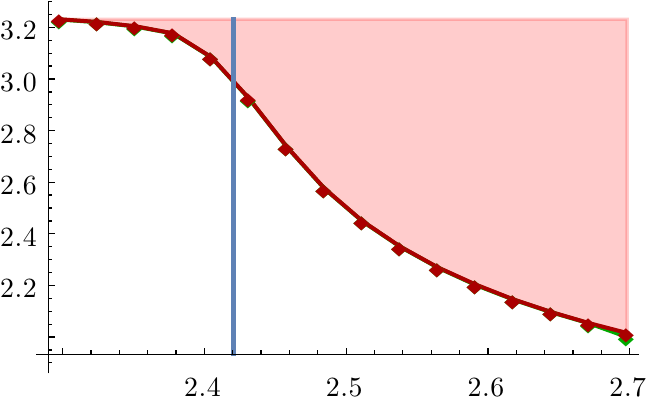}{$\Delta_{\mathrm{gap}}$}{$\lambda^2_{u\,u\,u^2}$}} \,
\subfloat[Theory $D_4$, $r=2$]{\includegraphicsWlabel[scale=1.03]{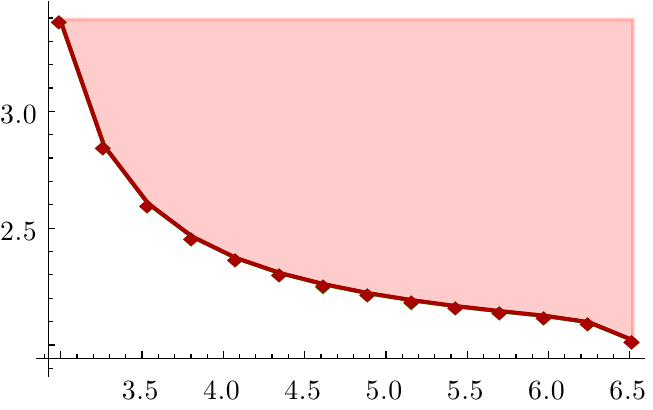}{$\Delta_{\mathrm{gap}}$}{$\lambda^2_{u\,u\,u^2}$}}

\caption{Upper bound on the dimension of the lightest scalar neutral unprotected operator $\Delta_{\mathrm{gap}}$ as a function of the OPE coefficient $\lambda_{u\,u\,u^2}^2$ for various rank-one theories. The upper area (shaded in red) is disallowed. The blue vertical line marks the value found by localization (see Table~\protect\ref{tSW}).}\label{fig:scans}
\end{figure}

In this subsection we present upper bounds on the dimension of the lightest unprotected neutral operator appearing in the OPE $\phi_r\times\phib_r$. The definition of the bootstrap problem to consider was given in subsection~\ref{sec:gap_bounds}. The following plots will present a scan on several values of the OPE coefficient $\lambda_{u\,u\,u^2}^2$ ranging between the bounds obtained in the previous subsection. The result is an upper bound on the lightest conformal dimension, which we denote as $\Delta_{\mathrm{gap}}$. It is interesting to notice that the upper end of the OPE window requires a very low $\Delta_{\mathrm{gap}}$, suggesting that probably the true value of $\lambda_{u\,u\,u^2}^2$ is located towards the lower end. This conclusion stems from the expectation that a strongly interacting theory is not expected to have operator dimensions too close to the free theory value. The plots are presented in Figure~\ref{fig:scans}.

\section{Large charge asymptotics}\label{Sec:LargeCharge}

The two-point functions $G_{nn}$ behave in a universal way as the charge (and, consequently, the dimension) of the operators is sent to infinity. More precisely, if we denote $nr\equiv\CJ$, we have the following asymptotic behavior~\cite{Hellerman:2017sur,Hellerman:2020sqj,Hellerman:2021duh}\footnote{After completion of the present paper, we learned that in \cite{Hellerman:2018xpi} this asymptotic formula has been improved, resumming the large-charge expansion. See also \cite{Beccaria:2020azj} for the higher-rank generalization.}
\eqn{
G_{nn}\;\underset{n\to\infty}{\sim}\;\tilde{\CY}\,\Gamma(\CJ+1) \left(\frac{N_\CO}{2\pi R}\right)^{2\CJ} \CJ^\alpha\,,
}[asympLargeN]
where $\tilde{\CY}$ and $N_\CO$ are constants and $\alpha$ is a constant related to the $a$-anomaly of the SCFT considered.\footnote{Not to be confused with the $\alpha$ of Table~\ref{tablead}.} The values $\alpha$ for the theories of interest are summarized in Table~\ref{tab:fits}.

\begin{figure}
\centering
\subfloat[{$r=\frac65$}]{\includegraphics[scale=1.4]{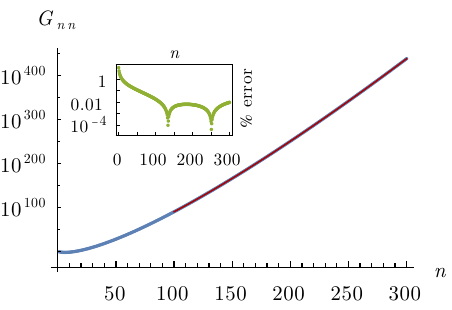}}
\subfloat[{$r=3$}]{\includegraphics[scale=1.4]{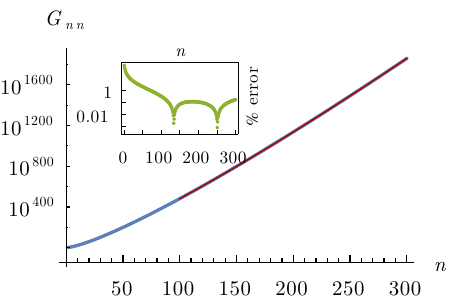}}
\caption{Comparison of $G_{nn}$ with the asymptotic form \asympLargeN where $c_{F}$ and $R$ have been set to $1$.}\label{fig:largeCharge}
\end{figure}

In this section we want to compare our results with this asymptotic expansion. This will provide a nontrivial check of our method and will also allow us to estimate the values of $\tilde{\CY}$ and $N_\CO$. The formula~\eqref{OPEcoeff2}, with $C_{ij}$ given by \eqref{cij2}, can in principle be computed exactly for all $n$, but since we want to study the behavior at very large $n$ it is easier to simply evaluate it numerically because this will significantly speed up the computation of the determinants. We found that the agreement with the formula~\asympLargeN is excellent already at $n\sim100$. In Figure~\ref{fig:largeCharge} we plot the values of $G_{nn}$ (in blue) together with the asymptotic form (in red). For brevity,  we only show $\mathcal{H}_0$ and MN with flavor group $E_6$. The other AD or MN theories look similar.. The small inset instead shows the percent error in green. Then in Table~\ref{tab:fits} we show our best estimates for $\tilde{\CY}$ and $N_\CO$.

\begin{table}
\centering
\setstretch{1.4}
\begin{tabular}{|c|ccccccc|}
\hline
$r$               & $\frac65$ & $\frac43$ & $\frac32$ & $2$ & $3$ & $4$ & $6$ \\
\hline
$\alpha$          & $\frac3{10}$ & $\frac12$ & $\frac34$ & $\frac32$ & $3$ & $\frac92$ & $\frac{15}2$ \\
$\sqrt{c_F}N_\CO$ & $2.197$      & $2.411$   &  $2.641$    &  $3.1416$   & $3.676$ &  $3.937$  &  $4.176$ \\
$\tilde\CY$       & $1.164$      & $1.111$   &  $1.010$    &  $0.6347$   & $0.137$ &  $1.619\times10^{-2}$  &  $6.530\times10^{-5}$ \\
\hline
\end{tabular}
\setstretch{1}
\caption{In the second row we show the values of the parameter $\alpha$ of \asympLargeN. In the last two rows we show the best fit estimates for $\tilde{\CY}$ and $N_\CO$ coming from the plots of Figure~\ref{fig:largeCharge}.}\label{tab:fits}
\end{table}

\section{Conclusions and Outlook}\label{Sec:Conclusions}
In this paper we have computed the OPE coefficients of some four-point correlators for $\mathcal{N}=2$ SCFT theories of the AD type. We have performed the computations using both localization and the bootstrap method as summarized in Table~\ref{tab:summary}. We find surprising agreement between the two methods, except for a few values. We remark that our results for the OPE coefficient $\lambda_{u\,u\,u^2}$ of the rank-one AD theories were also found in \cite{Grassi:2019txd}, extrapolating from the large-charge expansion, and in \cite{Cornagliotto:2017snu, Gimenez-Grau:2020jrx} using the bootstrap method.

\begin{table}[h]
\centering
\setstretch{1.05}
\noindent\parbox[t]{.46\textwidth}{%
\begin{tabular}[t]{|r|l|ccc|}
\hline
OPE & method & $\mathcal{H}_0$           & $\mathcal{H}_1$            & $\mathcal{H}_2$          
\\
\hline
\multirow{3}{*}{
$\lambda^2_{u\,u\,u^2}$} &
\multirow{2}{*}{Boot.$\;\Big\lbrace$} & 2.167 & 2.359 & 2.698 
\\
    &                                & 2.142 & 2.215 & 2.298 
\\ & Localiz. & 2.098 & 2.241 & 2.421
                                     \\ \hline
\multirow{3}{*}{
$\lambda^2_{u\,u^2\,u^3}$} &
\multirow{2}{*}{Boot.$\;\Big\lbrace$} & 3.637 & 4.445 &\\
   &                            & 3.192 & 3.217 &
                                      
\\ & Localiz. & 3.300 & 3.674 & 4.175 \\ \hline
\end{tabular}
}\hfill
\parbox[t]{.46\textwidth}{%
\begin{tabular}[t]{|r|l|cc|}
\hline
OPE & method & $(A_1,A_4)$ & $(A_1, A_5)$ \\
\hline

\multirow{3}{*}{
$\lambda^2_{u\,u\,u^2}$} &
\multirow{2}{*}{Boot.$\;\Big\lbrace$} & 2.102 & 2.231 \\
      &                         & 2.024 & 2.055
      
 \\ & Localiz. & 1.878 & 1.929 \\\hline

\multirow{3}{*}{
$\lambda^2_{u\,v\,uv}$} &
\multirow{2}{*}{Boot.$\;\Big\lbrace$} & 1.125 & 1.233 \\
       &                          & 0.981 & 0.960
       \\ & Localiz.  & 1.043 & 1.039 \\\hline

\multirow{3}{*}{
$\lambda^2_{v\,v\,v^2}$}  &
\multirow{2}{*}{Boot.$\;\Big\lbrace$} & 2.533 & 2.709 \\
		&					  & 2.181 & 2.195 
		
		\\ & Localiz.  & 2.231 & 2.203 \\ \hline
									 
\end{tabular}
}

\caption{Summary of the results from localization and numerical bootstrap, reproduced here for convenience and for easing the comparison between the two methods.\label{tab:summary}}
\end{table}

Probably the most mysterious point, at the conceptual level, is why the localization formula, though truncated at $\mathcal{F}_1$, gives so accurate values for the OPE coefficients. Indeed, lacking a parametrically-small quantity that regulates the genus expansion, higher-genus corrections are in principle not expected to give negligible contributions. While it would be very interesting to have a deeper understanding of this point, it would also be important to have a handle on the functions $\mathcal{F}_g$ with $g\geq2$ for the AD theories, to try and estimate their effect on the OPE coefficients. We hope to report on this matter in the near future.

On the more technical level, the localization formula for two-point correlators we used in this paper relies on knowledge of the SW prepotential $\mathcal{F}_0$, and hence of the SW periods. For $SU(N)$ gauge theories with fundamental matter of rank higher than one, explicit expressions for the SW periods of AD theories are, to our knowledge, not known. Here we tackle this problem, by considering directly the SW curve in the vicinity of
the conformal points and computing explicitly the SW periods for the rank-two AD appearing in the moduli space of pure $SU(5)$ and $SU(6)$ theories. 
It would  be nice to extend our computation to more theories of the AD type. Also, for higher-rank theories, we have been elusive regarding the integration contours of the matrix-model integral, and only paid attention to pick those which makes the integral over the CB parameters convergent. While this seems a reasonable guiding principle, further clarifications are needed on this point.

\acknowledgments
We are particularly grateful to R.~Poghossian for illuminating discussions on the periods of Riemann surface.
We are also grateful to A.~Grassi, J.~Minahan, D.~Orlando, J.~Russo, and L.~Tizzano for discussions. AM would like to thank G.~Fardelli and A.~Gimenez-Grau for discussions. The work of AB and AM is supported by Knut and Alice Wallenberg Foundation under grant KAW 2016.0129 and by VR grant 2018-04438. AM and AB would like to thank the INFN and University of Rome Tor Vergata for their hospitality.
The computations in this work were enabled by resources in project SNIC 2020/15-320 and SNIC 2021/22-500 provided by the Swedish National Infrastructure for Computing (SNIC) at UPPMAX, partially funded by the Swedish Research Council through grant agreement no. 2018-05973.


\begin{appendix}

\section{The Seiberg-Witten prepotential for \texorpdfstring{$\boldsymbol{SU(2)}$}{SU(2)} gauge theory}\label{SWSU(2)gauge}

\subsection{Elliptic geometry}
  
  In the quartic form, an elliptic geometry can be written as
 \be
 w^2=\prod_{i=1}^4 (x-e_i) =x^4+d_3 \, x^3 +d_2 \, x^2+d_1 \, x+d_0 \,.
 \ee
 The first period is 
 \be\label{Eq.w1}
 w_1 =  { {\rm i} \over \pi}   \int_{e_1}^{e_2} {dx \over   w } =   \int_0^1 {dz \over \pi\sqrt{e_{13}e_{24} \, z(1-z)(1-z \zeta)}}= { _{2} F_1(\ft12,\ft12,1,\zeta)\over  \sqrt{e_{13} e_{24} }}\,,
 \ee
 with $z={(x-e_1)e_{24}\over (x-e_4)e_{21}}$ and 
 \be
 \zeta ={e_{12} e_{34} \over e_{13}e_{24}  }\,.
 \ee
  The second period is obtained permuting the roots $(123)\to (231)$ leading to
   \be\label{Eq.w2}
 w_2 = {{\rm i} \over \pi }  \int_{e_2}^{e_3} {dx \over w } = { _{2} F_1\left(\ft12,\ft12,1, {e_{23} e_{14} \over  e_{21}e_{34}  }  \right)\over  \sqrt{e_{12} e_{43} }}= { {\rm i} \,  _{2} F_1(\ft12,\ft12,1,1-\zeta)  \over  \sqrt{e_{13} e_{24}}}\,,
 \ee
  where in the last equation we used the identity
 \be
 _{2} F_1\left(\ft12,\ft12,1,  x \right) =(1-x)^{-{1\over 2}}  {}_{2} F_1\left(\ft12,\ft12,1, {x\over x -1} \right)\,. \label{idk}
 \ee
     Ordering the roots
 \be
 e_1<e_2<e_3<e_4
 \ee
 one finds that $\zeta^{-1}>1$, so $\zeta^{-1} \notin [0,1]$ as required. 
   The torus complex structure is defined as the ratio of the periods
 \be
  \tau = {w_2\over w_1} ={{\rm i} \, {}_{2} F_1(\ft12,\ft12,1,1-\zeta)  \over _{2} F_1(\ft12,\ft12,1,\zeta)  }\,.
 \ee
    Alternatively, the periods of the curve can be expressed in terms of permutation invariant quantities  $D$, $\Delta$ given in terms of the expansion coefficients $d_n$ rather than the roots.  
   The two invariants are defined as
 \bea
 D&=&\ft{1}{16} \sum_{i\neq j\neq k \neq l} e_{ij}^2 \, e_{kl}^2=d_2^2-3\, d_1 \, d_3+12 d_0\nn\\
  \Delta&=&\prod_{i< j } e_{ij}^2 =d_2^2-3\, d_1 \, d_3+12 d_0=-27 d_1^4-4 d_3^3 d_1^3+18 d_2 d_3 d_1^3-4 d_2^3 d_1^2+d_2^2 d_3^2 d_1^2\,,\nn\\
  &&-6 d_0 d_3^2
   d_1^2+144 d_0 d_2 d_1^2+18 d_0 d_2 d_3^3 d_1-192 d_0^2 d_3 d_1-80 d_0 d_2^2 d_3
   d_1+16 d_0 d_2^4\nn\\
   &&-27 d_0^2 d_3^4+256 d_0^3-128 d_0^2 d_2^2-4 d_0 d_2^3 d_3^2+144
   d_0^2 d_2 d_3^2\,.\label{ddelta}
 \eea
  In terms of these variables, we can define the following two functions
  \bea
 w_3 &=& D^{-{1\over 4}} \,_{2} F_1\left(\ft{1}{12},\ft{5}{12},1,J^{-1}
 \right) \,,\label{f1f2}\\
 w_4 &=& {\rm i} D^{-{1\over 4}} \,\left[ -_{2} F_1\left(\ft{1}{12},\ft{5}{12},1,J^{-1}\right)
 +{\Gamma(\ft{5}{12})\Gamma(\ft{1}{12}) \over 2 \pi^{3\over 2} }  \,_{2} F_1\left(\ft{1}{12},\ft{5}{12},\ft12,1{-}J^{-1}
 \right) \right]\,, \nn
 \eea
 with
 \be
 J(\zeta)  ={4 D^3 \over 27\, \Delta } = {4 (1-\zeta+\zeta^2) \over 27 \zeta^2(1-\zeta)^2 }\,. 
 \ee

Using quadratic and cubic transformation identities of hypergeometric functions one can check that for  
  \bea\label{Weakw1w2}
 \zeta \ll1 : && \qquad   w_3=w_1   \quad, \quad w_4=w_2 \,. 
 \eea
 These formulae can be analytically continued away from $\zeta$ small,  where the periods $w_1$, $w_2$ can be written as linear combinations of $w_3$, $w_4$, related to those in the original region by modular transformations.
  
   The same formulae are obtained by writing the curve in the cubic Weiertrass form
   \be
 w^2=\prod_{i=1}^3 (x-e_i) = x^3+a_2 x+a_3\,,
 \ee
  where now
  \be
  \zeta= {e_{12}   \over e_{13}  } \, ,\ \qquad J= {4 a^3\over 4 a^3+27 b^2}\,.
  \ee
  
\subsection{SU(2) gauge theories with matter}

The curves of SU(2) gauge theories with fundamental matter of equal masses will be written as
\be
y^2+P(x) y+{\Lambda^{4-N_f}\over 4} \, (x-m)^{N_f}=0\,.
\ee
The branch cuts are located at
   \be
 w(x)^2=P(x)^2  -\Lambda^{4-N_f} \, (x-m)^{N_f} =\prod_{i=1}^4 (x-e_i)=\sum_{n=0}^4 d_n \, x^n \, ,
 \ee
 The function $P(x)$ for $N_f\leq 3$  is given by
 \be
  P(x)=
\left\{
\begin{array}{cc}
  x^2-u & N_f=0,1     \\
   x^2-u +{\Lambda^2\over 4}  & N_f=2     \\
    x^2-u +{\Lambda\over 4}  ( x- 3  m) & N_f=3  \,,   \\
\end{array}
\right.
  \ee
 We will compute the periods using the permutation invariant variables $D$ and $\Delta$, so we write
  \bea
{\partial a(u)  \over \partial u} &=& \oint_\alpha {\partial \lambda \over \partial u} = {1\over 2 \pi i}  \oint_{\alpha}  {dx\over w(x) }=w_1
\,,\nn  \\
{\partial a_D(u)  \over \partial u} &=& \oint_\alpha {\partial \lambda \over \partial u} =    {1\over 2 \pi i}  \oint_{\beta}  {dx\over w(x) } =w_2\,,
 \label{periods}
\eea
 with $w_1$, $w_2$ given by \eqref{Eq.w1} and \eqref{Eq.w2}. 
  The parameter $u$ is chosen such that the SW prepotential $\lambda$ behaves at large $x$ as
  \be
-2\pi {\rm i}  \lambda = x {d  \log y(x) \over dx} \approx \sum_{n=0}^\infty   { \left\langle   {\rm tr} \varphi^n  \right \rangle  \over x^n } \approx  2+{2u\over x^2}+\ldots  
  \ee
leading to $u= \ft12{\rm tr} \varphi^2$.  The weak coupling expansion of periods can alternatively be obtained via localization with instanton-counting parameter $q$ related to $\Lambda$ via
  \be 
 4q=\Lambda^{4-N_f}\,.
 \ee

\subsubsection[$N_f=0$ fundamentals]{$\boldsymbol{N_f=0}$ fundamentals}
Let us consider first the pure gauge theory. The SW curve becomes
   \be
  w(x)^2= (x^2-u)^2-\Lambda^4\,,
  \ee
  giving the invariants
  \bea
  D&=& 4(4u^2-3 \Lambda^4)\,, \nn\\
  \Delta &=& 256\Lambda^8(u^2- \Lambda^4)\,, \label{disc0}
 \eea
  Combining (\ref{periods}) with \eqref{Weakw1w2} and \eqref{f1f2}, one finds the weak coupling  expansion
  \bea
  {\partial a\over \partial u} &=& {1\over \sqrt{u}} \left( \frac{1}{2}+  {3\Lambda^4\over 32 u^2}+  {105\Lambda^8\over 2048 u^4} +\ldots \right)\,.
    \eea
   Integrating over $u$, inverting to find $u(a)$, and evaluating the second period one finds
   \bea
   a(u) &=& \sqrt{u}-{\Lambda^4\over 16 u^{3\over 2} } - {15 \Lambda^8\over 1024 u^{7\over 2} } -{105 \Lambda^{12} \over 16384 u^{11\over 2} } +\ldots\,,\nn\\
   u(a) &=& a^2 +  {\Lambda^4\over 2^3 a^2}+  {5\Lambda^8\over 2^9 a^6}+  {9 \Lambda^{12}\over 2^{12}\, a^{10} }+\ldots\,, \nn\\
   {\cal F}_0(a) &=& a^2  \log\left( 64 {a^4\over \Lambda^4} \right) -6 a^2 - {\Lambda^4\over 8 a^2}-  {5 \Lambda^8\over 1024 \,a^6}
   - {3\Lambda^{12}\over 4096 \,a^{10} } +\ldots\,. \label{au0}
   \eea
  The same formula follows from the Nekrasov partition function
    \be
    {\cal F}_{\rm loc}={-}\frac{2 q}{  \left(4 a^2{-}\epsilon^2 \right)}{-}\frac{q^2 \left(20 a^2{+}7 \epsilon _1^2{+}7 \epsilon
   _2^2{+}16 \epsilon _1 \epsilon _2\right)}{\left(4 a^2 {-}\epsilon^2\right){}^2  \left(4
   a^2{-}(2 \epsilon _1{+}\epsilon _2)^2\right)   \left(4 a^2{-}(\epsilon _1{+}2 \epsilon _2)^2\right) }+\ldots
   \ee
   sending $\epsilon_{1,2}$ to zero and setting $4q=\Lambda^4$.

    Using the expansions above one can check the relations \cite{Nakajima:2003uh}
\bea
u&=&\frac{\theta_3^4(q_{\rm IR})+\theta_2^4(q_{\rm IR})}{\theta_3^2(q_{\rm IR})\theta_2^2(q_{\rm IR})}\Lambda^2\,,\nonumber\\
\frac{\partial u}{\partial a}&=&\frac{\sqrt{2} \Lambda}{\theta_3(q_{\rm IR})\theta_2(q_{\rm IR})}\,,
\label{transprop}
\eea
where the $\theta_i(q_{\rm IR})$ are the standard theta functions and $q_{\rm IR}=e^{\pi {\rm i} \partial a_D/\partial a}$.

\subsubsection[$N_f=1$ fundamentals]{$\boldsymbol{N_f=1}$ fundamentals}

  Next we consider the $N_f=1$ theory.
  The curve becomes
  \be
  w(x)^2= \left(x^2-u \right)^2-\Lambda^3 (x-m)\,.
  \ee
 The invariants read
 \bea
 D &=&  4(4u^2+3 m \Lambda^3)\,,  \nn\\
 \Delta &=& \Lambda^6 \left(-27 \Lambda ^6+256 \Lambda^3 m^3+256 m^2 u^2-288 \Lambda ^3 m u-256 u^3\right) \,. \label{discnf1}
 \eea
 At weak coupling, combining (\ref{periods}) with \eqref{Weakw1w2} and \eqref{f1f2}, one finds
 \bea
 a(u) &=&\sqrt{u}+\frac{\Lambda ^3 m}{16 u^{3/2}}+\frac{3 \Lambda ^6 \left(u-5 m^2\right)}{1024 u^{7/2}}-\frac{35 \Lambda ^9 \left(m u-3 m^3\right)}{16384 u^{11/2}}+ \ldots\,,\nn\\
u(a) &=&a^2  -\frac{\Lambda ^3 m}{8 a^2}+\frac{\Lambda ^6 \left(5 m^2-3 a^2\right)}{512 a^6} +\frac{\Lambda ^9 \left(7 a^2 m-9 m^3\right)}{4096 a^{10}} {+}\ldots\,,\nn\\
{\cal F}_0(a) &=&  a^2  \log\left( -64{\rm i} {a^4\over \Lambda^4} \right) -{9 a^2\over 2}  -\ft12(a+m)^2\log\frac{(a+m)}{\Lambda}-\ft12(a-m)^2\log\frac{(a-m)}{\Lambda}\nn\\
&&+\frac{\Lambda ^3 m}{8 a^2}+\frac{\Lambda ^6 \left(3 a^2-5 m^2\right)}{1024 a^6}+\frac{\Lambda ^9 m \left(9 m^2-7 a^2\right)}{12288 a^{10}}+\ldots\,.  \label{au2}
\eea
 The same formula follows from the partition function computed using localization
 \be
{\cal F}_{\rm loc}=\frac{2 m q}{4a^2-\epsilon ^2}
+\frac{q^2 \left(3 \left(4 a^2-\epsilon ^2\right)^2-4 m^2 \left(20 a^2+7 \epsilon _1^2+7 \epsilon _2^2+16 \epsilon _1 \epsilon _2\right)\right)}{4 \left(4 a^2-\epsilon
   ^2\right)^2 \left(4 a^2-\left(2 \epsilon _1+\epsilon _2\right){}^2\right) \left(4 a^2-\left(\epsilon _1+2 \epsilon _2\right){}^2\right)} \ee
 sending $\epsilon_{1,2}$ to zero and setting $4q=\Lambda^3$. 

At strong coupling, near the AD conformal point $u_*={3 \Lambda^2 \over 4}$, $m_*=-{3\Lambda\over 4}$,  one finds
\be
\Delta \approx-432 \Lambda^8  (u-u_*)^2\, \qquad , \qquad  D=24 \Lambda^2 \, (u-u_*)  \qquad , \qquad J=-{128 (u-u_*)\over 27 \Lambda^2} \,.
\ee
 Expanding  (\ref{f1f2}) in this limit we get
 \be
 f_i\sim D^{-{1\over 4} } \, J^{1\over 12}\sim (u-u_*)^{-\frac{1}{6}}\,.
 \ee
 Integrating over $u$ one finds
\bea
a_D \sim a-a_* &\sim& (u-u_*)^{5\over 6}\qquad , \qquad \tau= {a_D\over a-a_*} =e^{\pi {\rm i} \over 3} 
\eea
 where we used the fact that $J=0$ for $\tau=e^{\pi {\rm i} \over 3}$ up to SL$(2,\mathbb{Z})$ transformations.

\subsubsection[$N_f=2$ fundamentals]{$\boldsymbol{N_f=2}$ fundamentals}

  Next we consider the $N_f=2$ theory with equal masses.
  The curve becomes
  \be
  w(x)^2= \left(x^2-u+ {\Lambda^2 \over 4} \right)^2-\Lambda^2(x-m)^2\,.
  \ee
 The invariants read
 \bea
 D &=& 16 u^2-12 m^2 \Lambda^2-4 u \Lambda^2+\Lambda^4 \,, \nn\\
 \Delta &=& 16  \Lambda^4(u^2-m^2 \Lambda^2)  (4u-4 m^2-\Lambda^2)^2\,.  \label{disc2}
 \eea
  At weak coupling, combining (\ref{periods}) with \eqref{Weakw1w2} and \eqref{f1f2}, one finds
 \bea
 a(u) &=&\sqrt{u}{-} \frac{\Lambda ^2 \left({u}+m^2\right)}{16 u^{3/2}}{-}\frac{3 \Lambda ^4 \left(5 m^4{+}2 m^2
   u{+}u^2\right)}{1024 u^{7/2}}{-}\frac{5 \Lambda ^6 \left(21 m^6{+}7 m^4 u{+}3 m^2 u^2{+}u^3\right)}{16384
   u^{11/2}}+\ldots\,,\nn\\
u(a) &=&  a^2{+}{a^2{+}m^2 \over 8 a^2}\Lambda^2+{a^4{-}6 a^2 m^2{+}5 m^4\over 512 a^6} \Lambda^4  {+} {5a^4{-}14 a^2 m^2{+}9 m^4\over 4096 a^{10}} m^2 \Lambda^6{+}\ldots\,,\nn\\
{\cal F}_0(a) &=&  a^2  \log\left( 64 {a^4\over \Lambda^4} \right) -3 a^2 -(a+m)^2\log\frac{(a+m)}{\Lambda}-(a-m)^2\log\frac{(a-m)}{\Lambda}\nn\\
&&{-}{\Lambda^2 (a^2{+}m^2) \over 8 a^2}{-}  {\Lambda^4 (5m^4{-}6  a^2 m^2{+}a^4)\over 1024 \,a^6}  \label{au2}
   +\ldots\,.  \label{au2}
\eea
 The same formula follows from the partition function computed using localization
 \be
{\cal F}_{\rm loc}= -\frac{q \left(4 \left(a^2{+}m^2\right){-}\epsilon ^2\right)}{2 \left(4 a^2{-}\epsilon ^2\right)}{-}\frac{q^2 \left(\left(4 a^2{-}\epsilon ^2\right)^2
   \left({-}4 a^2{+}24 m^2{+}\epsilon ^2\right){-}16 m^4 \left(20 a^2{+}7 \epsilon _1^2{+}7 \epsilon _2^2{+}16 \epsilon _1 \epsilon _2\right)\right)}{16
   \left(4 a^2{-}\epsilon ^2\right)^2 \left(4 a^2{-}\left(2 \epsilon _1{+}\epsilon _2\right){}^2\right) \left(4 a^2{-}\left(\epsilon _1{+}2 \epsilon
   _2\right){}^2\right)}
 \ee
sending $\epsilon_{1,2}$ to zero and setting $4q=\Lambda^2$.

At strong coupling, near the AD conformal point $u_*={\Lambda^2 \over 2}$, $m_*={\Lambda\over 2}$,  one finds
\be
\Delta \approx 256 \,\Lambda^6\, (u-u_*)^3 \, \qquad , \qquad  D=12 \Lambda^2 \, (u-u_*)  \qquad , \qquad J=1 \,.
\ee
 Expanding  (\ref{f1f2}) in this limit we get
 \be
 f_i\sim D^{-{1\over 4} } \, J^{1\over 12}\sim (u-u_*)^{-\frac{1}{4}}\,.
 \ee
 Integrating over $u$ one finds
\bea
 a_D\sim a-a_* &\sim& (u-u_*)^{3\over 4}\qquad , \qquad \tau= {a_D\over a-a_*} ={\rm i} 
\eea
 where we used the fact that $J=1$ for $\tau={\rm i}$ up to SL$(2,\mathbb{Z})$ transformations.

\subsubsection[$N_f=3$ fundamentals]{$\boldsymbol{N_f=3}$ fundamentals}

  Finally we consider the $N_f=3$ theory with equal masses.
  The curve becomes
  \be
  w(x)^2= \left(x^2-u+ {\Lambda \over 4} (x-3m) \right)^2-\Lambda(x-m)^3\,.
  \ee
 The invariants read
 \bea
 D &=&\frac{\Lambda ^4}{256}-\frac{3 \Lambda ^3 m}{8}+\frac{1}{2} \Lambda ^2 \left(9 m^2-2 u\right)+12 \Lambda  m \left(m^2+u\right)+16 u^2   \label{discnf3}\,, \\
 \Delta &=& \frac{\Lambda ^2}{8}  \left(2 m^2-\Lambda  m-2 u\right)^3 \left(256 \Lambda  m^3-3 \Lambda ^2 m^2+96 \Lambda  m u+256 u^2-\Lambda ^2 u\right) \,.\nn 
 \eea
  At weak coupling, combining (\ref{periods}) with \eqref{Weakw1w2} and \eqref{f1f2}, one finds
 \bea
 a(u) &=&\sqrt{u}+ \frac{\Lambda m \left(m^2+3  u\right)}{16 u^{3/2}}-\frac{\Lambda ^2 \left(15 m^6+27 m^4 u+21 m^2 u^2+u^3\right)}{1024 u^{7/2}} +\ldots\,,\nn\\
u(a) &=& a^2 -\frac{\Lambda m  \left(3 a^2 +m^2\right)}{8 a^2}+\frac{\Lambda ^2 \left(a^2-m^2\right)^2 \left(a^2+5 m^2\right)}{512 a^6} {+}\ldots\,,\nn\\
{\cal F}_0(a) &=&  a^2  \log\left( -64 {{\rm i} a^4\over \Lambda^4} \right) -{3 a^2\over 2} -\ft32(a+m)^2\log\frac{(a+m)}{\Lambda}-\ft32(a-m)^2\log\frac{(a-m)}{\Lambda}\nn\\
&&+\frac{\Lambda  m^3}{8 a^2}-\frac{\Lambda ^2m^2 \left(3 a^4 -9 a^2 m^2+5 m^4\right)}{1024 a^6} \label{au2nf3}
   +\ldots\,.  \label{au2}
\eea
 The same formula follows from the partition function computed using localization
  \be
{\cal F}_{\rm loc}= q \frac{ (3m-\epsilon) (4a^2-\epsilon^2)+4 m^3 }{2 \left(4 a^2{-}\epsilon ^2\right)} +\ldots
 \ee
sending $\epsilon_{1,2}$ to zero and setting $4q=\Lambda$. 

At strong coupling, near the AD conformal point $u_*={5 \Lambda^2 \over 64}$, $m_*=-{\Lambda\over 8}$,  one finds
\be
\Delta \approx-27 \Lambda^4  (u-u_*)^4\, \qquad , \qquad  D=16\, (u-u_*)^2  \qquad , \qquad J=-{16384 (u-u_*)^2\over 729 \Lambda^4} \,.
\ee
  Expanding  (\ref{f1f2}) in this limit we get
 \be
 f_i\sim D^{-{1\over 4} } \, J^{1\over 12}\sim (u-u_*)^{\frac{1}{3}}\,.
 \ee
 Integrating over $u$ one finds
\bea
 a_D\sim a-a_* &\sim& (u-u_*)^{2\over 3}\qquad , \qquad \tau= {a_D\over a-a_*} =e^{\pi {\rm i} \over 3}
\eea
 where we used the fact that $J=0$ for $\tau=e^{\pi {\rm i} \over 3}$ up to SL$(2,\mathbb{Z})$ transformations.

 \section{Conformal  block recursion relations}
 \label{appendixrecursion}
 
In equation~\eqref{eq:blockFactoriz} we mentioned that the derivatives of conformal blocks can be written as a positive factor times a polynomial. This property can be checked rather easily. First let us take the explicit expression of $g_{\Delta,\ell}$ in terms of hypergeometric functions from~\eqref{blockDef} with~\eqref{ksDef}. The derivatives with order higher than two acting on $k^s_\beta(z)$ can be recast as a first derivative or the function itself times polynomials, thanks to the hypergeometric equation. Then it suffices to do a Pade approximation of $k^s_\beta(1/2)$ and $\partial_zk^s_\beta(1/2)$. After this, it can be explicitly checked that the denominators will turn out to be positive for all unitary values of $\Delta$. See for example~\cite{Poland:2011ey}.

Let us now motivate the existence of such an approximation in general. Conformal blocks can be seen as meromorphic functions of $\Delta$ and representation theory can be used to predict their pole structure. Indeed their poles correspond to null states that arise in the degenerate Verma modules $\CV_{\Delta,\ell}$ when $\Delta$ assumes specific non-unitary values. The presence of these poles can be intuitively understood from the schematic definition of a conformal block as a projector onto a given conformal multiplet inserted inside a four-point function
\eqn{
\frac{g_{\Delta,\ell}(z,\zb)}{(x_{12}^2)^{\Delta_\phi}(x_{34}^2)^{\Delta_\phi}} = \sum_{n,m=0}^\infty\big\langle \phi(x_1)\phi(x_2)\big|\Psi_n^{\Delta,\ell}\big\rangle\,\left(\big\langle\Psi_n^{\Delta,\ell}\big|\Psi_m^{\Delta,\ell}\big\rangle\right)^{-1}\,\big\langle\Psi_m^{\Delta,\ell}\big|\phi(x_3)\phi(x_4)\big\rangle\,,
}[eq:cbSchematic]
where $|\Psi_0^{\Delta,\ell}\rangle$ is the state corresponding to the conformal primary of dimension $\Delta$ and spin $\ell$ and $n,m$ enumerate its descendants. The factor in the middle is the inverse of the matrix of two-point functions. If a certain Verma module $\CV_{\Delta,\ell}$ has a null descendant --- which can only happen for non-unitary values of $\Delta$ and $\ell$ --- the matrix of two-point functions will have a zero eigenvalue, thus leading to a pole in~\eqref{eq:cbSchematic}. These poles have been classified in~\cite{Penedones:2015aga}. The residue of each pole is a sum over all the descendants of the singular state, which means that it can be written as a conformal blocks with shifted parameters, times possibly a coefficient~$R_*$
\eqn{
g_{\Delta,\ell}(z,\zb) \quad\underset{\Delta\to\Delta^*}{\longrightarrow}\quad  \frac{R_*\, g_{\Delta^*,\ell^*}(z,\zb)}{\Delta - \Delta^*}\,.
}[]
With this property one can approximate the blocks very efficiently using a recursion relation. First we remove the essential singularity in $\Delta$ by defining a block $h_{\Delta,\ell}$ as follows
\eqn{
g_{\Delta,\ell}(z,\zb) = r^\Delta\,h_{\Delta,\ell}(z,\zb)\,,
}[]
where $r$ is related to the cross ratios $z$ and $\zb$ as follows 
\eqn{
r = \sqrt{|\rho\bar\rho|}\,,\qquad \rho = \frac{z}{(1+\sqrt{1-z})^2}\,,\qquad  \bar\rho = \frac{\zb}{(1+\sqrt{1-\zb})^2}\,.
}[]
Now $h_{\Delta,\ell}$, if regarded as a complex function of $\Delta$, is meromorphic and it approaches a constant at infinity, which means that it is fully specified by its poles
\eqn{
h_{\Delta,\ell}(z,\zb) = h_{\infty,\ell}(z,\zb) + \sum_{\mathrm{poles}\,\Delta_A^*} \frac{R_A\, r^{n_A}\,h_{\Delta^*_A,\ell^*_A}(z,\zb)}{\Delta - \Delta^*_A} + \sum_{\mathrm{double\,poles}\,\Delta_B^*} \frac{R_B\, r^{n_B}\,h_{\Delta^*_B,\ell^*_B}(z,\zb)}{(\Delta - \Delta^*_B)^2}\,.
}[eq:recRelBlocks]
The sum over double poles appears only for CFTs in even spacetime dimensions. Triple and higher poles never appear. The precise expressions for $h_{\infty,\ell}$, $R_A$, $\Delta_A^*$, $\ell_A^*$ and $n_A$ are given in~\cite{Penedones:2015aga}. The values of the poles $\Delta^*_B$ are also available in~\cite{Penedones:2015aga} (where it can be seen that two types of poles overlap) but the residues are unpublished. The detailed recursion relation is nevertheless implemented in an openly available software for all dimensions.\footnote{Available at \href{https://gitlab.com/bootstrapcollaboration/scalar_blocks}{\texttt{gitlab.com/bootstrapcollaboration/scalar\_blocks}}.} One can then truncate the sum by keeping only a finite number of poles. Noting that, since $n_A$ is always at least~$2$, the sum is suppressed by a factor of $r^2$, which, at the crossing symmetric point, is about~$0.029$. This allows us to solve the system iteratively and obtain a rapidly converging power series in $r$.
This type of recursion relation first appeared in the context of two-dimensional CFTs~\cite{Zamolodchikov:1984eqp, Zamolodchikov:1987eqp} but was later extended to general dimensions in~\cite{Kos:2013tga}. See also~\cite{Kos:2014bka} for a more detailed exposition.

Now that we have such a rational approximation of the conformal blocks, it is easy to take derivatives. Letting $r_* = 3-2\sqrt{2}$ being the value of $r$ at $z=\zb=1/2$ we have
\eqna{
\partial_z^n\partial_\zb^m\,g_{\Delta,\ell}\big(\tfrac12,\tfrac12\big) &= r_*^\Delta\Bigg(q^{m,n}_\ell(\Delta) +  \sum_{\mathrm{poles}\,\Delta_A^*} \frac{a_{\ell,A}^{m,n}}{\Delta-\Delta^*_A}\Bigg)\\
&= \frac{r_*^\Delta}{\prod_{\mathrm{poles}\,\Delta_A^*}(\Delta-\Delta_A^*)}\,p^{m,n}_\ell(\Delta)\,,
}[eq:derivApprox]
for some polynomial $q^{m,n}_\ell$ and some coefficients $a_{\ell,A}^{m,n}$. In the second line we simply took a common denominator. As we have explained earlier, the origin of the poles $\Delta_A^*$ guarantees that, for unitary values of $\Delta$, the denominator is always a positive function.

In order to have very precise expressions for the blocks, one would like to keep many poles. If the polynomials are too big, however, the semidefinite programming problem becomes more expensive. A good trade-off between precision and performance is obtained by computing the blocks at very high recursion order and then Padé approximating the obtained expression to a rational function with fewer poles. This is known as the pole-shifting procedure and it was introduced in~\cite{Kos:2013tga}.

 \section{Crossing equations of the system of mixed correlators}\label{app:mixed_crossing_eqns}

In this appendix we show the crossing equations for the system of correlators involving two distinct chiral fields $\phi_{r_1}$, $\phi_{r_2}$ and their antichiral partners. Before writing the crossing vectors we need to introduce the blocks for unequal external operators. In order not to clutter the equations with too many subscripts, we depart from the notation in the main text and use $g$ in place of $\CG^s$, $\tilde\CG$ in place of $\CG^t$ and $\CG$ in place of $\CG^u$. This is also consistent with previous works~\cite{Lemos:2015awa}. With this notation, the generalizations of~\eqref{blockDef} read
\eqna{
g^{(r,r')}_{\Delta,\ell}(z,\zb) & =\frac{z\zb}{z-\zb}\bigl(k^s_{\Delta+\ell}(z)k^s_{\Delta-\ell-2}(\zb)- z\leftrightarrow\zb\bigr)\,,\\
\CG^{(r)}_{\Delta,\ell}(z,\zb) &= \frac{z\zb}{z-\zb}\bigl(k^u_{\Delta+\ell}(z)k^u_{\Delta-\ell-2}(\zb)- z\leftrightarrow\zb\bigr)\,,\\
\tilde{\CG}^{(r)}_{\Delta,\ell}(z,\zb)& =\frac{z\zb}{z-\zb}\bigl(k^t_{\Delta+\ell}(z)k^t_{\Delta-\ell-2}(\zb)- z\leftrightarrow\zb\bigr)\,,\\
}[blockDefMixed]
where now the $k$ functions are defined as
\eqna{
k^s_\beta(z) &= 
z^{\frac{\beta}2}\,{}_2F_1\lnsp\mleft(\frac{\beta-r}2,\frac{\beta+r'}2;\beta;z\mright)\,, \\
k^u_\beta(z) &= 
z^{\frac{\beta}2}\,{}_2F_1\lnsp\mleft(\frac{\beta-r}2,\frac{\beta-r+4}2;\beta+2;z\mright)\,,\\
k^t_\beta(z) &= 
z^{\frac{\beta}2}\,{}_2F_1\lnsp\mleft(\frac{\beta-r}2,\frac{\beta+r}2;\beta+2;z\mright)\,.
}[kDefMixed]
Next we construct the crossing functions which generalize~\eqref{Ffunctions}
\threeseqn{
F^{ijkl}_{\pm,\Delta,\ell}(z,\zb) &= v^{\frac{r_j+r_k}2} g^{(r_{ij},r_{kl})}_{\Delta,\ell}(z,\zb)  \pm u^{\frac{r_j+r_k}2} g^{(r_{ij},r_{kl})}_{\Delta,\ell}(1-z,1-\zb)\,,
}[]{
\CF^{ijkl}_{\pm,\Delta,\ell}(z,\zb) &= v^{\frac{r_j+r_k}2} \CG^{(r_{ij})}_{\Delta,\ell}(z,\zb)  \pm u^{\frac{r_j+r_k}2} \CG^{(r_{ij})}_{\Delta,\ell}(1-z,1-\zb)\,,
}[]{
\tilde{\CF}^{ijkl}_{\pm,\Delta,\ell}(z,\zb) &= v^{\frac{r_j+r_k}2} \tilde{\CG}^{(r_{ij})}_{\Delta,\ell}(z,\zb)  \pm u^{\frac{r_j+r_k}2} \tilde{\CG}^{(r_{ij})}_{\Delta,\ell}(1-z,1-\zb)\,.
}[][]
with $r_{ij} = r_i - r_j$ and $r_i$ is both the R-charge and the dimension of $\phi_{r_i}$. The blocks $\CG$ and $\tilde{\CG}$ have only one superscript because they will only appear with $r_{ij}=- r_{kl}$ and $r_{ij}=r_{kl}$ respectively.

Let us abbreviate $\phi_i = \phi_{r_i}$ and $\phib_i=\phib_{r_i}$. The crossing equation can then be written as follows
\eqna{
&\sum_{\CO\in\phi_1\times\phi_2} |\lambda_{\phi_1\phi_2\CO}|^2\, \vec{V}^{r_1+r_2}_{\Delta,\ell} + \sum_{i=1,2} \sum_{\CO\in\phi_i\phi_i} |\lambda_{\phi_i\phi_i\CO}|^2 \, \vec{V}^{2r_i}_{\Delta,\ell}
\\&+
\sum_{\CO\in\phi_1\times\phib_2} |\lambda_{\phi_1\phib_2\CO}|^2\, \vec{V}^{r_1-r_2}_{\Delta,\ell}+
\sum_{\CO\in\phi_i\times\phib_i} (\lambda_{\phi_1\phib_1\CO}^*\;\lambda_{\phi_2\phib_2\CO}^*)\cdot\vec{\mathcal{V}}^{\mathrm{neutral}}_{\Delta,\ell} \cdot
\begin{pmatrix}
\lambda_{\phi_1\phib_1\CO}\\
\lambda_{\phi_2\phib_2\CO}
\end{pmatrix} = \\&=
-\vec{V}_\unit - \frac{1}{6\llsp c}\vec{V}_T\,,
}[]
with the crossing vectors $\vec{V}^{q}_{\Delta,\ell},\vec{V}^\mathrm{neutral}_{0,0},\vec{V}_{T}$ and the crossing matrix $\vec{\mathcal{V}}^{\mathrm{neutral}}_{\Delta,\ell}$ defined below
%
\newcommand{\FF}[1]{F_{\pm,\Delta,\ell}^{#1}}
\newcommand{\CFF}[1]{\mathcal{F}_{\pm,\Delta,\ell}^{#1}}
\newcommand{\CtFF}[1]{\tilde{\mathcal{F}}_{\pm,\Delta,\ell}^{#1}}
\newcommand{\CFFm}[1]{\mathcal{F}_{-,\Delta,\ell}^{#1}}
\newcommand{\CtFFm}[1]{\tilde{\mathcal{F}}_{-,\Delta,\ell}^{#1}}
\newcommand{\monel}{(-1)^\ell}
\fourseqn{\small
\vec{V}^{r_1+r_2}_{\Delta,\ell} = \left[
\begin{array}{c}
\mp\monel\FF{2121} \\
0_2 \\ 
\mp \FF{1221} \\
0_6
\end{array}
\right]\,,\qquad
\vec{V}^{2r_1}_{\Delta,\ell} = \left[
\begin{array}{c}
0_6 \\
\mp\monel\FF{1111} \\
0_4
\end{array}
\right]\,,
}[]{
\vec{V}^{2r_2}_{\Delta,\ell} = \left[
\begin{array}{c}
0_9 \\
\mp\monel\FF{2222} \\
0
\end{array}
\right]\,,\qquad
\vec{V}^{r_1-r_2}_{\Delta,\ell} = \left[
\begin{array}{c}
\monel\CtFF{2121} \\ 
\CFF{1221} \\
0_8
\end{array}
\right]\,,
}[]{
\vec{\mathcal{V}}^{\mathrm{neutral}}_{\Delta,\ell} = \left[
\begin{array}{c}
\boldsymbol0_2\\
\begin{pmatrix}
0 & \mp\frac12\CFF{1122} \\ \mp\frac12\CFF{1122} & 0
\end{pmatrix}\\
\begin{pmatrix}
0 & \frac\monel2\CtFF{1122} \\ \frac\monel2\CtFF{1122} & 0
\end{pmatrix}\\
\begin{pmatrix}
\monel\CtFF{1111} & 0 \\ 0 & 0
\end{pmatrix}\\
\begin{pmatrix}
\CFFm{1111} & 0 \\ 0 & 0
\end{pmatrix}\\
\begin{pmatrix}
0 & 0 \\ 0 & \monel\CtFF{2222}
\end{pmatrix}\\
\begin{pmatrix}
0 & 0 \\ 0 & \CFFm{2222}
\end{pmatrix}\\
\end{array}
\right]\,,\hspace{3.8em}
}[]{
\vec{V}_{\unit} = \left[
\begin{array}{c}
0_2 \\ 
\mp\mathcal{F}^{1122}_{\pm,0,0} \\
\tilde{\mathcal{F}}^{1122}_{\pm,0,0} \\
\tilde{\mathcal{F}}^{1111}_{\pm,0,0} \\
\mathcal{F}^{1111}_{-,0,0} \\
\tilde{\mathcal{F}}^{2222}_{\pm,0,0} \\
\mathcal{F}^{2222}_{-,0,0}
\end{array}
\right]\,,\qquad
\vec{V}_{T} = \left[
\begin{array}{c}
0_2 \\ 
\mp r_1r_2\,\mathcal{F}^{1122}_{\pm,2,0} \\
r_1r_2\,\tilde{\mathcal{F}}^{1122}_{\pm,2,0} \\
r_1^2\,\tilde{\mathcal{F}}^{1111}_{\pm,2,0} \\
r_1^2\,\mathcal{F}^{1111}_{-,2,0} \\
r_2^2\,\tilde{\mathcal{F}}^{2222}_{\pm,2,0} \\
r_2^2\,\mathcal{F}^{2222}_{-,0,0}
\end{array}
\right]\,.\hspace{2.2em}
}[][mixedVecDef]
In the above vectors we defined as $0_n$ a sequence of $n$ zeros and as $\boldsymbol0_n$ a sequence of $n$ null two-by-two matrices. Entries with a $\pm$ span two rows, one for each choice of sign.

For the allowed values of $\Delta$ and $\ell$ in each OPE refer to section~\ref{sec:multiplets}. The vector $\vec{V}_\unit$ represents the contribution of the identity and the vector $\vec{V}_{T}$ represents the contribution of the stress tensor multiplet, namely $A_2\overbar{A}_2[0;0]_2^{(0;0)}$. They are both particular cases of $\vec{\mathcal{V}}^{\mathrm{neutral}}_{\Delta,\ell}$ dotted with the appropriate OPE coefficients.

\section{Normalization of the operators}

In this appendix, we show the operator normalization that we chose for out computations. This will make the meaning of $\lambda_{ijk}$ unambiguous. Scalar operators are normalized as
\eqn{
\langle \CO_i(x_1) \COb_j(x_2)\rangle = \frac{\delta_{ij}}{(x_{12}^2)^{\Delta_i}}\,.
}[twopnorm]
Their OPE is fixed by defining the conformal blocks. Namely, we expand a four-point function as
\eqn{\langle \CO_1(x_1)\cdots\CO_4(x_4)\rangle = \frac{\left(\frac{x_{24}^2}{x_{14}^2}\right)^{\frac12\Delta_{12}}\left(\frac{x_{14}^2}{x_{13}^2}\right)^{\frac12\Delta_{34}}}{(x_{12}^2)^{\frac12(\Delta_1+\Delta_2)}(x_{34}^2)^{\frac12(\Delta_3+\Delta_4)}}\,\sum_{\Delta,\ell}\lambda_{12(\Delta,\ell)}\lambda_{34(\Delta,\ell)}^*\, g^{\Delta_{12},\Delta_{34}}_{\Delta,\ell}(z,\zb)\,,
}[]
where the function $g^{\Delta_{12},\Delta_{34}}_{\Delta,\ell}(z,\zb)$ is defined in~\eqref{blockDefMixed}. A property that this block satisfies is
\eqn{
g^{\Delta_{12},\Delta_{34}}_{\Delta,\ell}(z,z) \;\overset{z\to 0}{\longrightarrow}\; z^\Delta\lsp(\ell+1)\,.
}[gnorm]
The choices~\twopnorm and~\gnorm are enough to unambiguously determine the meaning of $\lambda_{ijk}$.

\end{appendix}

\bibliographystyle{JHEP}
\bibliography{refs}
\end{document}